\documentclass[conference]{IEEEtran}
\IEEEoverridecommandlockouts
\usepackage{cite}
\usepackage{amsmath,amssymb,amsfonts}
\usepackage{algorithmic}
\usepackage{graphicx}
\usepackage{textcomp}
\usepackage{amsmath}
\usepackage[english]{babel}
\usepackage{enumerate}
\usepackage{amsthm}
\usepackage{xcolor,colortbl}
\usepackage{graphicx, multicol}
\usepackage{subfig}

\theoremstyle{definition}

\newtheorem*{assumption*}{\assumptionnumber}
\providecommand{\assumptionnumber}{}
\makeatletter
\newenvironment{assumption}[1]
 {%
  \renewcommand{\assumptionnumber}{Assumption #1}%
  \begin{assumption*}%
  \protected@edef\@currentlabel{#1}%
 }
 {%
  \end{assumption*}
 }
\makeatother

\usepackage{xcolor}
\def\BibTeX{{\rm B\kern-.05em{\sc i\kern-.025em b}\kern-.08em
    T\kern-.1667em\lower.7ex\hbox{E}\kern-.125emX}}

\usepackage{tikz}
\newcommand*\circled[1]{\tikz[baseline=(char.base)]{
            \node[shape=circle,draw,inner sep=0.8pt] (char) {#1};}}

\newcommand{\algoname}{SybilWall} 
\newcommand{\labelflip}{label-flipping attack}
\newcommand{\backdoor}{backdoor attack}
\newcommand{\dense}{\textit{dense Sybil poisoning attack}}
\newcommand{\distributed}{\textit{distributed Sybil poisoning attack}}
\newcommand{\sparse}{\textit{sparse Sybil poisoning attack}}
\newcommand{\Dense}{\textit{Dense Sybil poisoning attack}}
\newcommand{\Distributed}{\textit{Distributed Sybil poisoning attack}}
\newcommand{\Sparse}{\textit{Sparse Sybil poisoning attack}}

\IEEEaftertitletext{\vspace{-1\baselineskip}\noindent\makebox[\linewidth][c]{\normalsize \textbf{ --- MSc. Thesis ---}}\vspace{1.5\baselineskip}}

\begin{document}

\title{Towards Sybil Resilience in Decentralized Learning
}

\author{\IEEEauthorblockN{Thomas Werthenbach}
\IEEEauthorblockA{
\textit{Delft University of Technology}\\
Delft, The Netherlands \\
T.A.K.Werthenbach@student.tudelft.nl}
\and
\IEEEauthorblockN{Johan Pouwelse}
\IEEEauthorblockA{
\textit{Delft University of Technology}\\
Delft, The Netherlands \\
J.A.Pouwelse@tudelft.nl}
}

\maketitle
\thispagestyle{plain}
\pagestyle{plain}

\begin{abstract}

Federated learning is a privacy-enforcing machine learning technology but suffers from limited scalability. This limitation mostly originates from the internet connection and memory capacity of the central parameter server, and the complexity of the model aggregation function. 
Decentralized learning has recently been emerging as a promising alternative to federated learning.
This novel technology eliminates the need for a central parameter server by decentralizing the model aggregation across all participating nodes.
Numerous studies have been conducted on improving the resilience of federated learning against poisoning and Sybil attacks, whereas the resilience of decentralized learning remains largely unstudied. 
This research gap serves as the main motivator for this study, in which our objective is to improve the Sybil poisoning resilience of decentralized learning.



We present \algoname{}, an innovative algorithm focused on increasing the resilience of decentralized learning against targeted Sybil poisoning attacks.
By combining a Sybil-resistant aggregation function based on similarity between Sybils with a novel probabilistic gossiping mechanism, we establish a new benchmark for scalable, Sybil-resilient decentralized learning.


A comprehensive empirical evaluation demonstrated that \algoname{} outperforms existing state-of-the-art solutions designed for federated learning scenarios and is the only algorithm to obtain consistent accuracy over a range of adversarial attack scenarios. We also found \algoname{} to diminish the utility of creating many Sybils, as our evaluations demonstrate a higher success rate among adversaries employing fewer Sybils.
Finally, we suggest a number of possible improvements to \algoname{} and highlight promising future research directions.
\end{abstract}


\section{Introduction}
The rise of machine learning has resulted in an increasing number of everyday-life intelligent applications. 
As such, machine learning has been used in personal assistants \cite{personalassistant}, cybersecurity \cite{cybersecurity}, and recommendations on social media \cite{mlalgosocialmedia} and music \cite{mlalgomusic}. 
However, accurate machine learning models require large training datasets \cite{moredataisbetter, moredataisbetter2}, which can be difficult to collect due to privacy concerns \cite{shao2019stochastic} and recent privacy legislation \cite{privacylegislationinml}. 
\textit{Federated learning} \cite{federatedlearning} has become a promising option for distributed machine learning. It has been proposed for the training of numerous industrial machine learning models \cite{automotionfl, automotionfl2, gboard, gboard2, nextwordprediction}. 
Moreover, federated learning ensures the protection of privacy, as the user's data will not leave their device.

In contrast to centralized machine learning, model training in federated learning takes place on end-users' personal devices, which are often referred to as \textit{edge devices} or \textit{nodes}. 
The resulting trained models are communicated to a central server, commonly referred to as the \textit{parameter server}, which aggregates these trained models into a single global model. 
By only sharing the end-user-trained models with the parameter server, the user's privacy is preserved, while obtaining comparable performance compared to centralized machine learning \cite{cheng2020federated}. 
Although there exist attacks in which training data can be reconstructed based on model gradients \cite{datareconstructionattack, 10003066}, defense mechanisms against this attack have been proposed \cite{electronics12020260, hashagainstreconstruction}. 

\begin{figure}
\hfill
  \subfloat[\label{fig:fl}Federated learning]
  {\includegraphics[height=3.5cm]{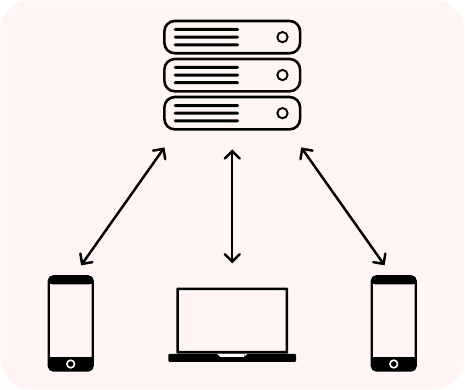}}
  \hfill
  \subfloat[\label{fig:dl}Decentralized learning]{\includegraphics[height=3.5cm]{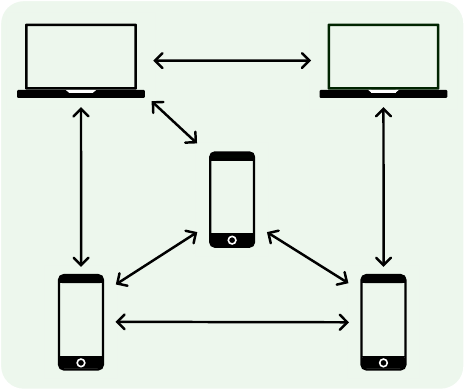}}
  \hfill
  \caption{\label{fig:flvsdl}Exemplary network topologies in federated learning and decentralized learning.}
\end{figure}

However, federated learning suffers from some disadvantages. 
First, the parameter server downloads the models of all participating nodes and broadcasts the aggregated global model each training round, inducing high communication costs and a potential bottleneck in the learning process. This may affect the overall convergence time \cite{communicationbottleneck}. 
Second, the scalability of the chosen aggregation function in terms of the number of nodes may vary greatly. In robust and secure federated learning aggregation methods, the incorporation of additional nodes during aggregation can result in a significantly increased computational effort for the parameter server \cite{scalabilitydepends}. 
Third, the parameter server performing the aggregation poses a single point of failure \cite{singlepointoffailure, decentralizedlearningvulnerable}. 
Disruptions to the parameter server can cause downtime and hinder the overall model training process, particularly when nodes require the globally aggregated model before proceeding the training. 
An upcoming alternative that aims to resolve these issues is \textit{decentralized learning} \cite{dl1, dl2, dl3, modest}, also commonly referred to as \textit{decentralized federated learning}. 
In decentralized learning, there exists no dedicated parameter server performing the aggregation, and the nodes form a distributed network, e.g., a peer-to-peer network. Each node in this network individually performs the aggregation using their neighbors' models (Figure \ref{fig:flvsdl}). 
This alleviates the scalability limitations and single point of failure issues imposed on federated learning and paves the path for boundless scalability.
While the information available during aggregation is more limited relative to federated learning, decentralized learning has the potential to obtain similar results compared to federated learning \cite{DLworks}. 


Although decentralized learning resolves the drawbacks faced by federated learning, it is still vulnerable to malicious environments \cite{decentralizedlearningvulnerable}. Since the predefined aggregation method in decentralized learning does not have access to all models in the network, aggregation is performed with less information compared to federated learning. This causes decentralized learning to have relatively lower resistance against possible poisoning attacks \cite{poisoningattack}. 
Poisoning attacks can generally be classified in two categories, namely those of \textit{targeted poisoning attacks} and \textit{untargeted poisoning attacks}. 
Targeted poisoning attacks focus on a specific goal that an adversary aims to achieve, while untargeted poisoning attacks aim to hinder the result of the training process without any particular goal in mind.
The effect of these attacks can often be amplified by combining them with a Sybil attack \cite{sybilattack}, in which an adversary creates a substantial number of virtual nodes to increase its influence. As such, an adversary may deploy the Sybil attack to spread their poisoned model more rapidly through the network. In this work, we focus exclusively on targeted poisoning attacks amplified by Sybil attacks in decentralized learning.

Prior work on resilience against Sybil poisoning attacks in distributed machine learning has mainly been done in federated learning settings. One popular example of such work is \textit{FoolsGold} \cite{foolsgold}, which aims to increase resilience against targeted Sybil poisoning attacks under the assumption that all Sybils will exhibit highly similar behavior. Experimental results suggest that FoolsGold can provide effective protection against Sybil attacks in small-scale federated learning.

In this work, we experimentally demonstrate \textit{FoolsGold}'s inability to scale to an unbounded number of nodes in federated learning and inept defensive capabilities against targeted poisoning attacks in decentralized learning. 

We suggest an improved version of FoolsGold, named \algoname{}, which shows significant resilience towards defending against targeted poisoning attacks while enjoying the boundless scalability offered by decentralized learning. More specifically, we achieve this by introducing a probabilistic gossiping mechanism for data dissemination. We performed an empirical evaluation of \algoname{} and found that it achieved satisfactory accuracy, convergence rate, and Sybil poisoning resilience on 4 different datasets. Moreover, comparative evaluations demonstrate \algoname{}'s superior Sybil resilience over numerous existing solutions.
Lastly, we found that \algoname{} successfully diminishes the utility of creating many Sybils.

To the best of our knowledge, this work is the first to propose a defensive algorithm against poisoning attacks amplified by the Sybil attack in decentralized learning. In short, our contributions are the following:
\begin{itemize}
    \item We define the Spread Sybil Poisoning attack in Section \ref{sec:threatmodelandassumptions} for effective Sybil poisoning attacks in decentralized learning and decompose it into three distinct scenarios.
    \item We present \algoname{}, a pioneering algorithm for Sybil poisoning resilience with boundless scalability in decentralized learning, in Section \ref{sec:design}.
    \item We performed an empirical evaluation of the performance of \algoname{} in Section \ref{sec:evaluation} on various datasets and against competitive alternatives.
\end{itemize}

\section{Background}
Federated learning was initially proposed by Google \cite{federatedlearning} as a means of training machine learning models on real user data without compromising user privacy. However, federated learning is associated with limitations in scalability. Decentralized learning is a promising alternative as it resolves scalability limitations through decentralization. Both distributed machine learning technologies are prone to poisoning attacks, of which the effects can be amplified by employing the Sybil attack.

\subsection{Federated learning}
Federated learning achieves privacy-enforcing machine learning by training all machine learning models on the edge devices (nodes) of the participating users, containing real user data (Figure \ref{fig:fl}). Training proceeds in synchronous rounds. During each training round, the participating nodes train the globally shared model on their private data for a predefined number of epochs and send the trained models to a central parameter server. The role of the parameter server is to aggregate all trained models into a global model without the need for the training data. After aggregation, the parameter server communicates the global model to all nodes, immediately followed by the start of the next training round. Alternatively, nodes may send \textit{gradients} to the parameter server rather than the trained model. These gradients are the result of training the model on the node's local dataset, e.g. with stochastic gradient descent (SGD), in that particular training round. These gradients can be used to compute the trained model by adding the gradients to the aggregated global model.

The original federated learning paper \cite{federatedlearning} suggests the usage of \textit{FedAvg}, which adopts a weighted average function as the aggregation function, such that the next global model $w^{t+1}$ is calculated as follows:
\begin{equation}
w^{t+1} = \sum_{i\in N} \frac{|\mathcal{D}_i|}{|\mathcal{D}|} w^t_i
\end{equation}
where $w^t_i$ is the model of node $i$ in round $t$, $N$ is the set of nodes, $\mathcal{D}_i$ corresponds to node $i$'s local dataset and ${\mathcal{D}}$ is the global distributed collection of data, such that $\mathcal{D} = \bigcup_{j \in N}\mathcal{D}_j$. 

The goal of the training process is to minimize the global loss function such that the global model $x$ approaches the optimal model $x^*$. More formally, the search for a global optimal model can approximately be defined as:
\begin{equation}
w^* = \arg\min_{w} \sum_{i \in N}\frac{|\mathcal{D}_i|}{|\mathcal{D}|} \mathcal{L}_i(w)
\end{equation}
where $\mathcal{L}_i$ is a node's loss function,  e.g. cross-entropy loss or negative log likelihood loss, using the node $i$'s local dataset.

In federated learning, all participating nodes are only connected to the parameter server, such that the network graph $\mathcal{G}$ is defined as a tuple of nodes and undirected edges $\langle N, E \rangle$, for which there exists a one-to-one mapping $N \to E$, such that for parameter server $s$, $\forall n \in N, \langle n, s \rangle \in E$.

\subsection{Decentralized learning}
\label{sec:dl}
\begin{figure}
    \centering
    \includegraphics[width=1\linewidth]{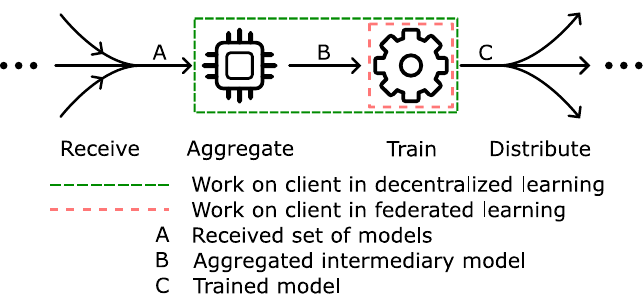}
    \caption{The general train-aggregate loop executed by all nodes participating in decentralized learning. The difference between the work performed by nodes in federated learning and decentralized learning is highlighted.}
    \label{fig:trainaggregateloop}
\end{figure}

Decentralized learning is an upcoming alternative to federated learning \cite{dl1, dl2, dl3, modest} (Figure \ref{fig:dl}). 
In contrast to federated learning, which relies on a parameter server to aggregate locally trained models, aggregation in decentralized learning takes place on a smaller scale, as it is performed by every participating node.
During each training round, every node trains their local aggregated model on their private data for a predefined number of epochs and then broadcasts only the trained model to its neighboring nodes. After receiving the trained models from all neighbors, every node aggregates both its own and its neighbors' trained models to construct the aggregated intermediary model, which is then used to train on during the next training round. This train-aggregate loop is depicted in Figure \ref{fig:trainaggregateloop}. 

By eliminating the need for a central parameter server, decentralized learning does not suffer from the scalability limitations encountered in federated learning. These improvements in scalability can be decomposed into three distinct aspects:

\begin{enumerate}
    \item \textit{Communication costs}: In federated learning, all models are downloaded and uploaded by the parameter server every training round, forming a communication bottleneck bounded by the parameter server's internet connection. Such bottlenecks are reduced in decentralized learning depending on a node's number of neighbors.
    \item \textit{Memory}: Storing all models in memory during aggregation may result in substantial memory usage. In decentralized learning, aggregation coincides with a significantly reduced number of models compared to federated learning, thus diminishing memory-related limitations.
    \item \textit{Aggregation time}: The time complexity of the more sophisticated aggregation functions may not scale linearly with respect to the number of participating nodes. Due to the decentralization of the aggregation, the number of models in each aggregation is greatly reduced.
\end{enumerate}

Furthermore, nodes likely do not have access to all information in the network. This may cause an impaired convergence rate compared to federated learning \cite{modest} and lead to increased vulnerability to potential attacks \cite{decentralizedlearningvulnerable}. However, decentralized learning has been shown to obtain comparable performance results compared to federated learning \cite{DLworks} and may sporadically outperform federated learning altogether \cite{dloutperformsdfl, dloutperformsdfl2}. 

A network graph $\mathcal{G}$ in decentralized learning can be defined as the tuple $\langle N, E \rangle$, where $N$ and $E$ represent the set of nodes and the set of undirected edges, respectively. The set of edges $E$ is restricted such that $\forall \langle n, m \rangle \in E$ it holds that $n \in N \wedge m \in N \wedge n \neq m$. We say that nodes $n$ and $m$ are neighbors if $\langle n, m \rangle \in E$. If $\langle n, m \rangle \notin E$, $n$ is a distant node (or indirect neighbor) in $m$'s perspective and vice versa. Moreover, the distance between any two nodes is defined as the minimum number of edges connecting the two nodes. In contrast to federated learning, the network graph $\mathcal{G}$ in decentralized learning does not contain a parameter server. 


\subsection{Targeted poisoning attacks}
A \textit{poisoning attack} is a type of Byzantine attack, which encapsulates all methods by which an adversary may attempt to compromise the integrity of the global model in decentralized learning and federated learning. Poisoning attacks typically include \textit{data poisoning} or \textit{model poisoning}. \textit{Data poisoning} involves malicious alteration of the training data, while \textit{model poisoning} entails adapting the process in which the model is trained \cite{poisoningattack}. In this work, we only consider data poisoning. Furthermore, poisoning attacks can be subdivided into two categories: \textit{targeted poisoning attacks} and \textit{untargeted poisoning attacks}. With untargeted poisoning attacks, the adversary aims to decrease the performance metric of the global model without any particular goal in mind. On the other hand, targeted poisoning attacks are employed to achieve a specific goal by manipulating the global model to behave in a deterministic manner that deviates from objectively correct behavior. In this paper, we exclusively consider two types of targeted poisoning attacks: the \labelflip{} \cite{labelflippingattack, labelflippingattack2} and the \backdoor{} \cite{backdoorattack, backdoorattack2, backdoorattack3}.

The \labelflip{} can be deployed as an attempt to increase the probability of two targeted classes being misclassified. More specifically, given two target classes $t_1$ and $t_2$, the aim of the \labelflip{} is to manipulate the model such that an arbitrary sample belonging to class $t_1$ is more likely to be classified as class $t_2$ by the global model and vice versa. One logical way of achieving this through data poisoning is by explicitly transforming the adversary's local dataset $\mathcal{D}$ into an adversarial dataset $\mathcal{D}'$ and train the adversarial model on this dataset. Given two target classes $t_1$ and $t_2$, this transformation can be defined as:

\begin{equation}
\label{eq:label_flip}
\begin{split}
\mathcal{D}' =\ & \{(x, y) \in \mathcal{D}\ |\ y \neq t_1 \wedge y \neq t_2\} \\ 
& \cup \{(x, t_1)\ |\ (x, y) \in \mathcal{D},\ y = t_2\} \\
& \cup \{(x, t_2)\ |\ (x, y) \in \mathcal{D},\ y = t_1\}    
\end{split}
\end{equation}

The \backdoor{} requires a more sophisticated manipulation of the training data. The objective of a \backdoor{} is to alter the global model such that any sample containing a specific predefined pattern is misclassified to a chosen target class. In the domain of image classification, this adversarial pattern could, for instance, correspond to a small square or triangle in the top left corner of the input image \cite{DBLP:journals/corr/abs-2011-01767}. Given a target class $t$ and a function $f$ that introduces a hidden pattern to input samples, the transformation applied on the adversary's local dataset $\mathcal{D}$ can be defined as:

\begin{equation}
\label{eq:backdoor}
\mathcal{D}' = \{(f(x), t)\ |\ (x, y) \in \mathcal{D}\}
\end{equation}

\subsection{The Sybil attack}
The Sybil attack \cite{sybilattack} is an adversarial strategy in distributed environments in which the attacker exploits the inability to verify the authenticity of any node's identity. Through the effortless creation of fake nodes, \textit{Sybils}, and strategical edges to honest nodes, the attacker may gain significantly more influence compared to honest nodes. We denote the edges between Sybils and honest nodes as \textit{attack edges}. A typical scenario in which the Sybil attack may be deployed is \textit{majority voting} \cite{levine2006survey, tran2009sybil}. In such a case, an attacker can trivially generate sufficient Sybils to outnumber all honest voters.

Methods for mitigating the Sybil attack through an admission control system to the decentralized network have been proposed \cite{4215910, souche, 4806900}, but are often not frictionless or are based on an invite-only system. Adoption of such systems may take place at a slower rate due to its decreased accessibility and usability \cite{captchalternatives}. The importance of frictionless admission becomes increasingly apparent considering that decentralized learning can be implemented as a background task \cite{gboard2}.

A network graph on which a Sybil attack is deployed can be defined as $\mathcal{G} = \langle N', E' \rangle$, such that $N' = N \cup \mathcal{S}$, where $\mathcal{S}$ is the unbounded set of Sybils created by the adversary. Note that Sybils and honest nodes are indistinguishable from the typical point of view. The modified set of edges $E'$ is defined as $E' = E\ \cup\ E_\mathcal{S}$, where $E_\mathcal{S}$ is the set of attack edges and edges between Sybils, which is highly dependent on the strategy of the adversary. Note that attack edges always consist of at least one Sybil, such that $\forall \langle i, j \rangle \in E_\mathcal{S}, i \in \mathcal{S} \vee j \in \mathcal{S}$.

In this work, we consider the targeted Sybil poisoning attack, in which an adversary aims to amplify the effects of a targeted poisoning attack by creating Sybils. These Sybils help spread the adversary's malicious model more rapidly and effectively throughout the network.

\section{Related work}
\label{sec:related_work}
Numerous studies have been conducted in order to improve poisoning resilience in a form of distributed machine learning. This section provides an overview of two existing defense mechanisms used in the empirical evaluation of \algoname{}.

\subsection{FoolsGold}
\label{sec:foolsgold}
FoolsGold \cite{foolsgold} is an algorithm designed to mitigate targeted Sybil poisoning attacks in federated learning settings. It builds on the assumption that Sybil model gradients show a substantially higher degree of similarity relative to that of honest model gradients, as they collaborate to reach the goal of a targeted poisoning attack. By computing the similarity between these model gradients, FoolsGold manages to successfully mitigate Sybil poisoning attacks in federated learning.

During aggregation, the parameter server first computes the pairwise cosine similarity score for all gradient histories. The gradient history of node $i$ in round $T$ is defined as $h^T_i = \sum_{t = 0}^T g_i^t$, where $g_i^t$ are the gradients of a model obtained by training the model on node $i$ in round $t$. However, as honest nodes may still produce similar gradient histories, this may result in an increased number of false positives. In an effort to decrease the number of false positives, FoolsGold implements a \textit{pardoning} mechanism. This pardoning mechanism multiplies each similarity score $s_{ij}$ by the ratio of the maximum score of node $i$ and the maximum score of node $j$ in the cases where the latter is greater, such that $s_{ij}$ is multiplied by $\frac{\max_v s_{iv}}{\max_v s_{jv}}$ if $\max_v s_{iv} < \max_v s_{jv}$.

Subsequently, the scores are aggregated for each node by taking the complement of the maximum score, such that node $i$'s aggregated score $s_i'$ can be defined as $s_i' = 1 - \max_v s_{iv}$. These aggregated scores now represent the extent to which a node can be trusted, based on its cosine similarity score. The aggregated scores are then rescaled such that the highest aggregated score equals $1$, as FoolsGold assumes the existence of at least one honest node. 
Each score now indicates a node's similarity to any other node, with a value close to 0 indicating high similarity, while a value near 1 suggests little similarity. These scores are then transformed using a bounded logit function to prioritize higher-scoring nodes. Finally, the scores are normalized and adopted as weights in a weighted average on the trained models to compute the aggregated model.

\begin{figure}
    \centering
        \includegraphics[width=0.485\linewidth]{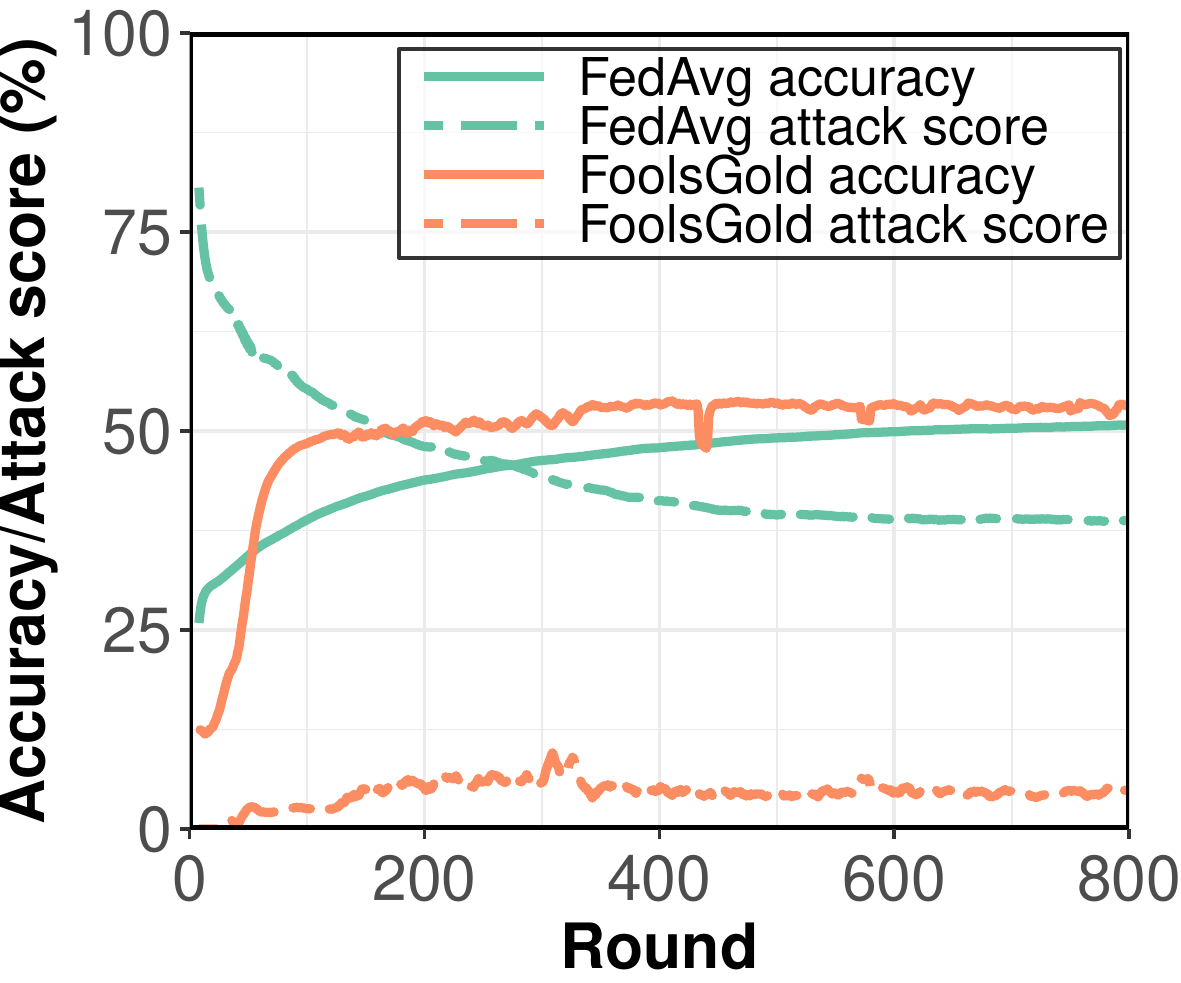}
  \caption{\label{fig:foolsgold-vs-fedavg-cifar-10-fl}FoolsGold and FedAvg in federated learning setting using the CIFAR-10 \cite{cifar10} dataset on a LeNet-5 \cite{lenet5} model.}
\end{figure}

\begin{figure}
    \centering    \includegraphics[width=0.485\linewidth]{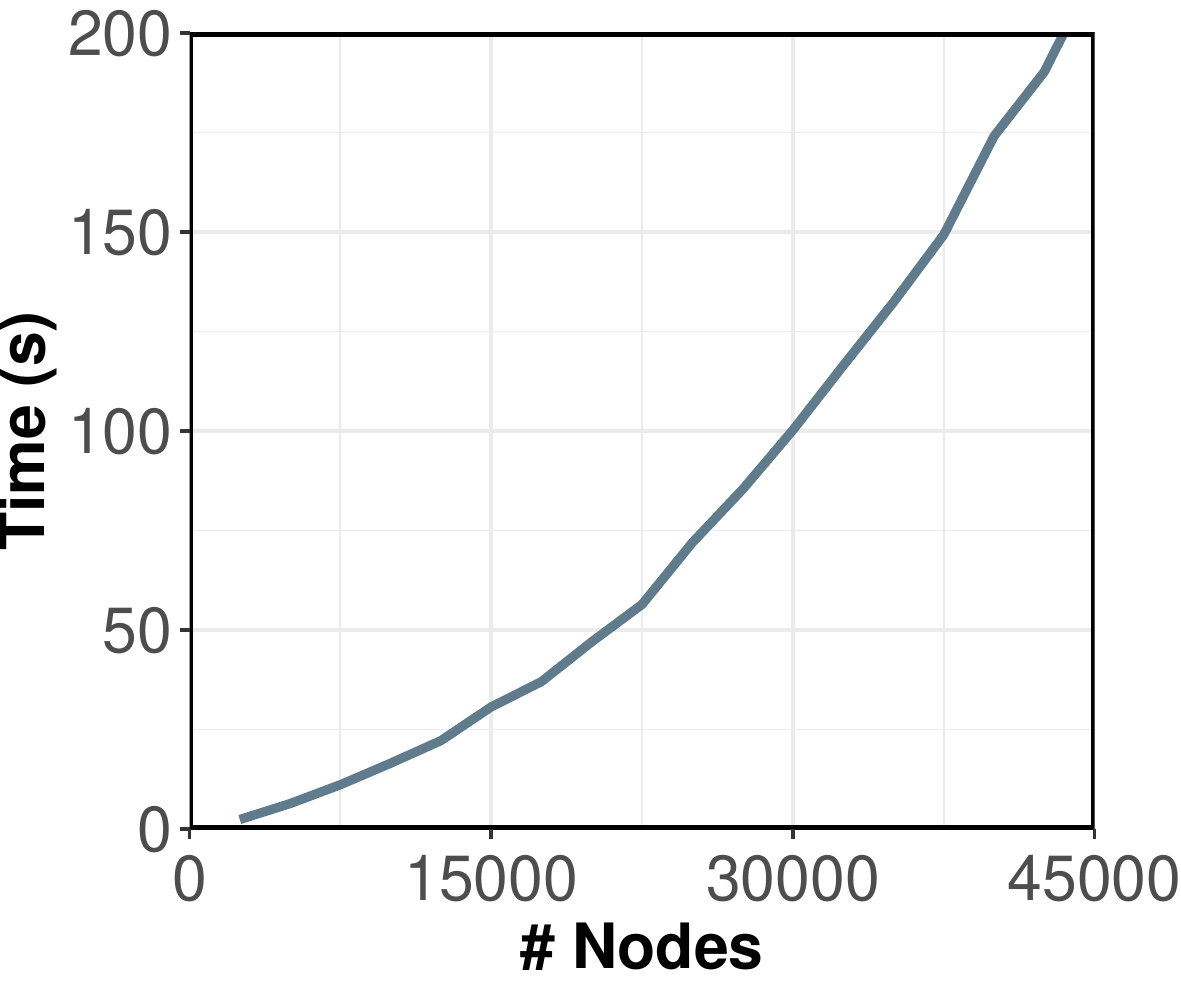}
    \caption{Pairwise cosine similarity computation time against the number of nodes (LeNet-5 \cite{lenet5}).}
    \label{fig:cosine_time}
\end{figure}

\begin{figure}[t]
\captionsetup[subfloat]{farskip=0pt,captionskip=1pt}
\centering
  \subfloat[\label{fig:foolsgold-our-algo-1}Network topology 1]{
       \includegraphics[width=0.485\linewidth]{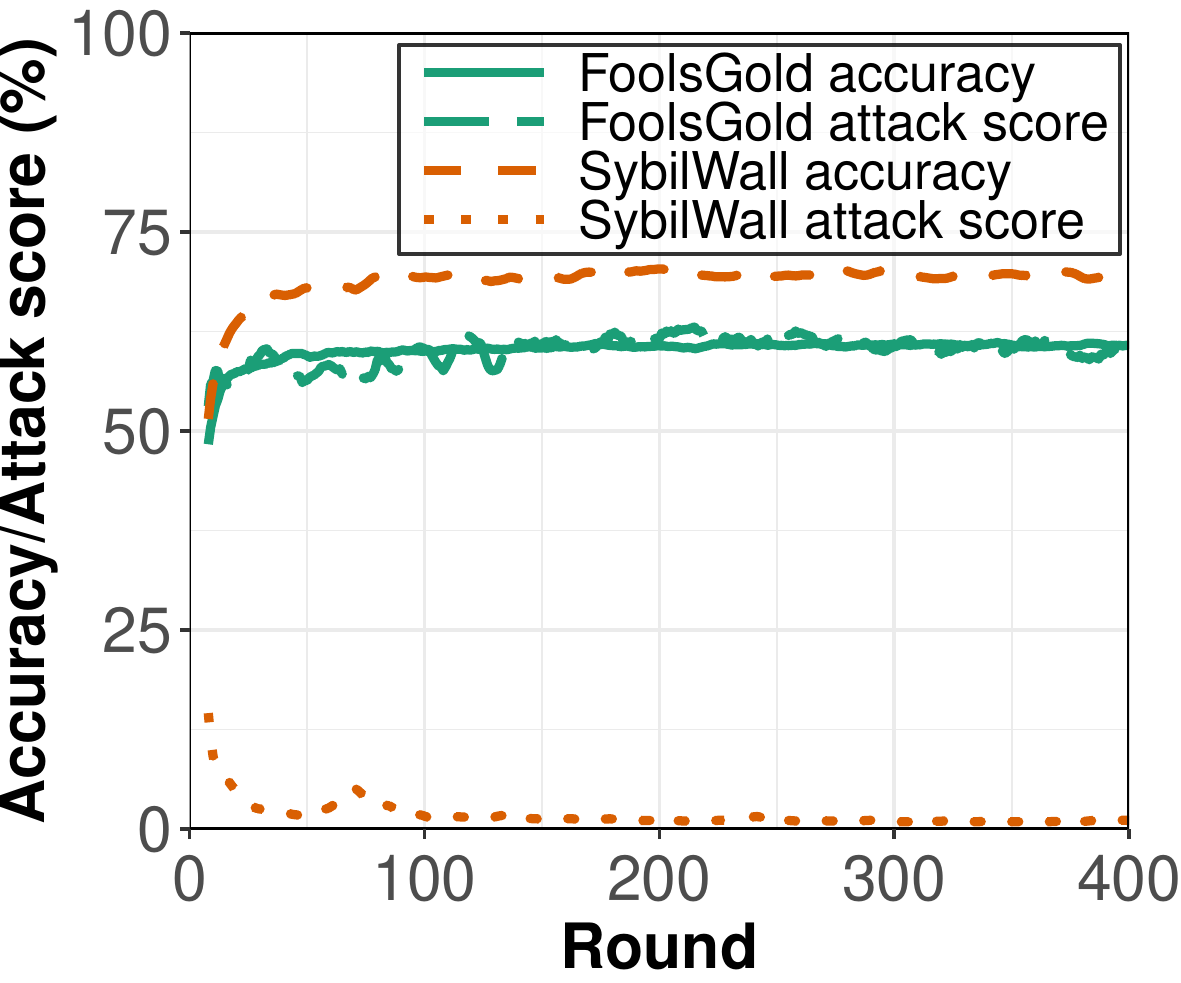}}
    \hfill
  \subfloat[\label{fig:foolsgold-our-algo-2}Network topology 2]{
        \includegraphics[width=0.485\linewidth]{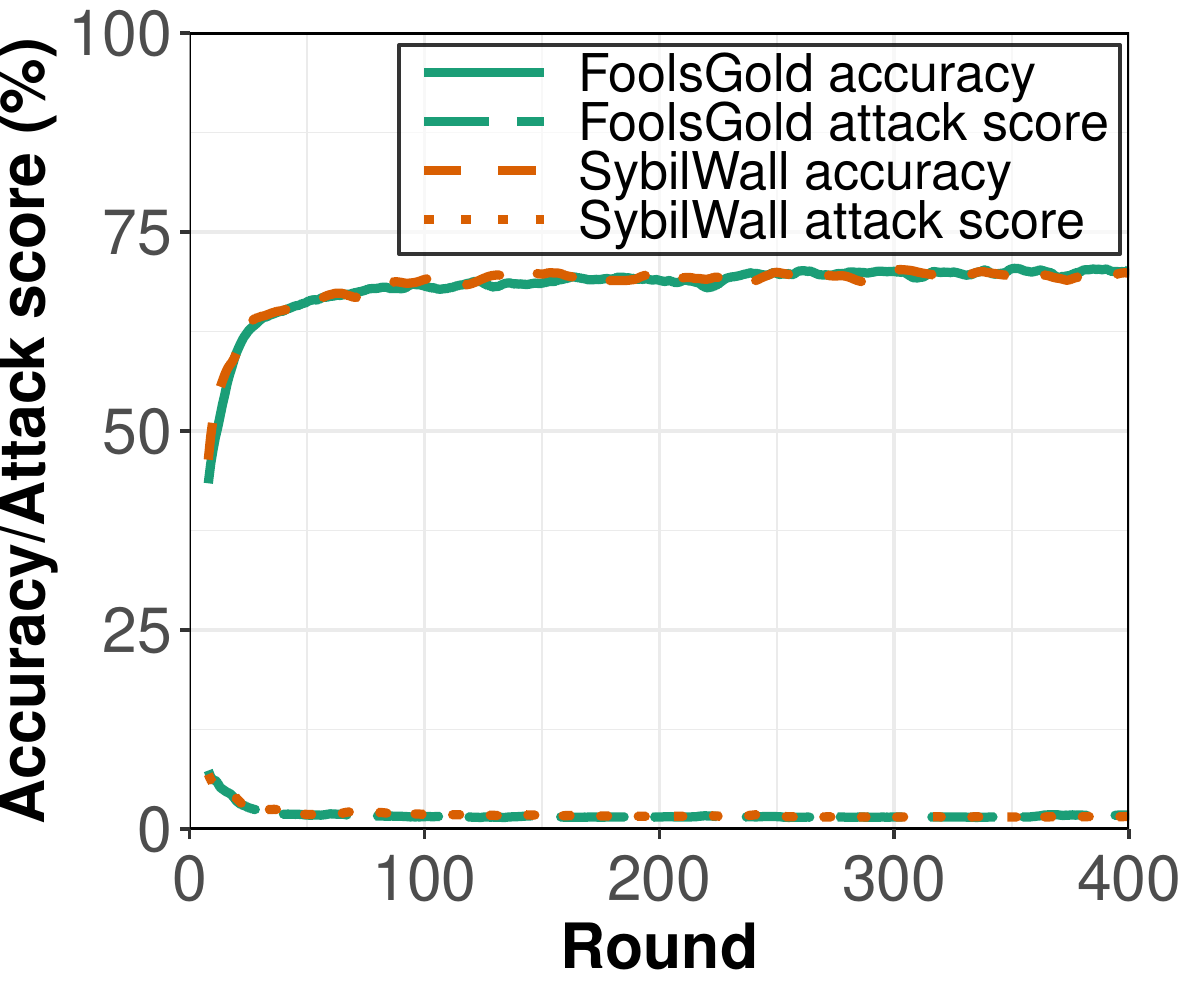}}
  \caption{\label{fig:foolsgold-vs-our-algo-fashionmnist-dl}FoolsGold and \algoname{} in decentralized learning using the FashionMNIST \cite{fashionmnist} dataset on a single-layer softmax neural network.}
\end{figure}

A reproduction of FoolsGold's results can be found in Figure \ref{fig:foolsgold-vs-fedavg-cifar-10-fl}, where the attack score represents the extent to which the attack was successful, e.g., the percentage of labels that are successfully flipped in the \labelflip{}. It becomes clear that FoolsGold shows significantly higher Sybil resilience compared to FedAvg. However, as discussed in Section \ref{sec:dl}, federated learning can be considered unscalable as the number of participating nodes increases. The limited scalability of FoolsGold is further highlighted by its $\mathcal{O}(n^2)$ pairwise cosine similarity computation and the memory capacity required to store these models. Figure \ref{fig:cosine_time} demonstrates the $\mathcal{O}(n^2)$ time complexity of the pairwise cosine similarity computation on the LeNet-5 model \cite{lenet5}. We further note that the experiments required to generate Figure \ref{fig:cosine_time} consumed the maximum memory allocated. This emphasizes the memory limitation associated with federated learning. Furthermore, Figure \ref{fig:foolsgold-vs-our-algo-fashionmnist-dl} shows the performance of FoolsGold in a decentralized setting against the performance of our improved solution, \algoname{}, based on FoolsGold's intuitions. When comparing both Figures \ref{fig:foolsgold-our-algo-1} and \ref{fig:foolsgold-our-algo-2}, it becomes clear that FoolsGold's performance heavily depends on the network topology, while \algoname{} demonstrates relatively higher and more consistent Sybil resilience.

\subsection{Krum}
Krum \cite{krum} attempts to improve the overall Byzantine resilience in distributed machine learning. This approach operates on the assumption that Byzantine model gradients are prone to deviate from the gradients produced by honest nodes. More specifically, the aggregation involves computing a score $s(w)$ for every received model $w$. This score corresponds to the sum of the squared distances between $i$ and its $n - f - 2$ nearest neighbours, where $f$ corresponds to the maximum number of Byzantine nodes Krum is configured to protect against. Finally, the model $m$ with the lowest score, such that $m = \arg\min_w s(w)$, is chosen as the next global model.



\section{Threat model and assumptions}
\label{sec:threatmodelandassumptions}
This section provides an overview of the assumptions and threat model used throughout this work.
\subsection{Adversarial assumptions}
\begin{assumption}{1}\label{as:limited_to_dl_api}
\textit{The adversary can only communicate with other nodes through the default decentralized learning API.}
\end{assumption}
As the adversary can only communicate with other nodes through the default decentralized learning API, it does not possess the ability to directly manipulate other nodes' local models or data. However, adversaries are not restricted in manipulating their own model, which is sent to neighbors. We also assume that the decentralized learning API enforces homogeneous model broadcasting. That is, every node broadcasts the same model to each of its neighbors every training round. In practice, this can be enforced by adopting existing algorithms \cite{contrib}. 
Lastly, we assume that the default decentralized learning API adopts the use of signatures to prevent spoofing. 

\begin{assumption}{2}\label{as:cryptographic_primitives_safe}
\textit{All used cryptographic primitives are secure.}
\end{assumption}
The signatures used by the decentralized learning API, as well as any other cryptographic primitives employed throughout this work, are assumed to be secure. 

\begin{assumption}{3}\label{as:adversary_not_constrained}
\textit{The adversary is unrestricted in both the quantity of Sybil nodes it can create and the selection of honest nodes it can form attack edges to.}
\end{assumption}

\begin{assumption}{4}\label{as:sybils_similar_updates}
\textit{Sybil models show high similarity compared to honest models.}
\end{assumption}

Given the context of targeted poisoning attacks, Sybils are created by an adversary to achieve a specific goal during decentralized learning. As these Sybils share their training dataset, their trained models will likely show a high similarity.

In contrast to prior work \cite{foolsgold}, we assume a high similarity between the trained models of Sybils, rather than the model gradients, i.e. the difference between the aggregated intermediary model and the trained model. Due to the lack of knowledge of the aggregated intermediary model between the aggregation and training stages (Figure \ref{fig:trainaggregateloop}), no node can ascertain the model gradients of another node in decentralized learning. 

\begin{assumption}{5}\label{as:no_additional_computation_for_new_sybils}
\textit{The creation of Sybils by the adversary does not increase its adversarial computing capabilities.}
\end{assumption}
Following Assumption \ref{as:sybils_similar_updates} and the lack of knowledge of the aggregated intermediary model, we must assume that each Sybil utilizes the same aggregated intermediary model. This assumption is enforced through Assumption \ref{as:no_additional_computation_for_new_sybils}, that is, the adversary does not have sufficient adversarial computing capabilities to execute the train-aggregate loop for each Sybil each round.

\subsection{Network restrictions}
\begin{assumption}{6}\label{as:degree_bounded}
  \textit{$\exists\ e \in \mathbb{N}$ such that $d_i \leq e$, $\forall i \in N$, where $d_i$ represents the degree of node $i$.}
\end{assumption}
We restrict the impact that any individual node may exercise on the network, by assuming existence of an upper bound on the degree of any node. Such bounds may arise naturally due to internet connection speeds, but may also be detected through existing algorithms. For example, a network latency-based avoidance mechanism \cite{latencyavoidance} can be used to discover multiple edges of a node. Another alternative is to perform a random walk or a breadth-first search, which are known to be biased toward high-degree nodes \cite{5462078}.

\begin{assumption}{7}\label{as:at_least_one_honest_neighbour}
  \textit{Every node has at least one honest neighbour.}
\end{assumption}
This final assumption is inherited from prior work \cite{foolsgold}, as the cosine similarity function requires a baseline for \textit{honest} work for measuring relative similarity. This might be achieved through an invite-only network with accountability \cite{souche}. We note that Eclipse attacks \cite{eclipseattack} are out of the scope of this work.

\subsection{Adversarial strategy}
\label{sec:adverserial_attack_strategy}
We define an intuitive and effective type of worst-case attack in similarity-based aggregation techniques in decentralized learning as \textit{Spread Sybil Poisoning Attacks} (SSP attacks). That is, the adversary aims to avoid detection by maximizing the distance between its attack edges. At the same time, the adversary attempts to increase the influence of the attack by minimizing the distance between any honest node and the nearest attack edge. The latter part of this problem resembles the \textit{Maximal Covering Location Problem} \cite{church1974maximal}, which is known to be an NP-Hard problem \cite{maximumnp-hard}. To determine the attack edge positions for SSP attacks, we propose a heuristic approach using the unsupervised clustering algorithm K-medoids \cite{kmedoids}, assigning attack edges to the medoids. 

Furthermore, we define a parameter for SSP attacks, $\phi$, which represents the average density of attack edges per node. Note that the attack edges are as spread out as possible, such that $\forall a_i, a_j \in \mathcal{A}, |a_i - a_j| \leq 1$, where $\mathcal{A}$ represents the set of the number of attack edges per node. For any value of $\phi$, each honest node receives $\lfloor \phi \rfloor$ or $\lceil \phi \rceil$ attack edges. Therefore, the total number of attack edges is denoted as $\lceil |N| \cdot \phi \rceil$. The remainder, defined by $\phi \bmod 1$, is distributed according to the K-medoids clustering algorithm. The resulting attack edge positions are then grouped and distributed over the Sybils while maintaining Assumption \ref{as:degree_bounded}. We define three attack scenarios for specific ranges of $\phi$. These attack scenarios are the following:
\begin{enumerate}[i.]
    \item \Dense{}. $\phi \geq 2$. Every honest node has at least two attack edges, whereas any distinct Sybil cannot form more than one attack edge to any given node. As a result, each honest node is a direct neighbor of at least two distinct Sybils.
    \item \Distributed{}. $\epsilon < \phi < 2$. There exists at least one node which is connected to fewer than 2 attack edges and will therefore only be connected to at most one Sybil.
    \item \Sparse{}. $\phi \leq \epsilon$. A low $\phi$ will result in sparse and distant attack edges. Any node has a probability of $\phi$ of being directly connected to a Sybil.
\end{enumerate}

\begin{figure*}
    \centering
    \includegraphics[width=1\linewidth]{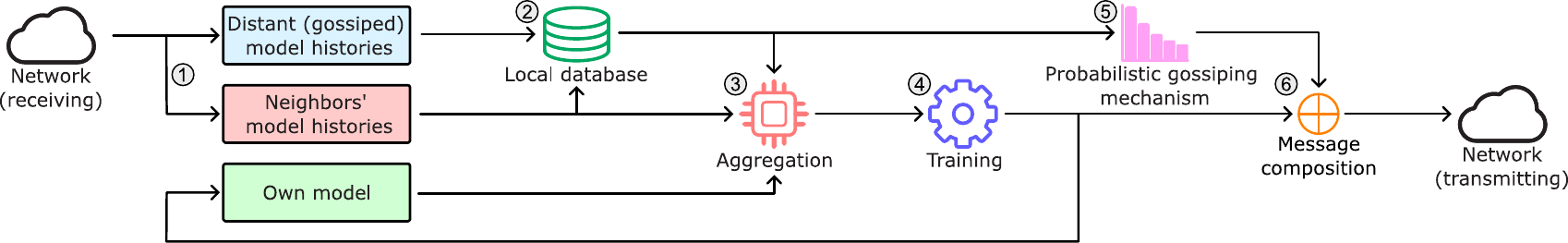}
    \caption{The Sybilwall architecture.}
    \label{fig:architecture}
\end{figure*}

\section{Design of \algoname{}}
\label{sec:design}
Our proposed algorithm, \algoname{}, takes inspiration from federated learning but was meticulously designed to enable boundless scalability through decentralization. Another primary purpose of SybilWall is the mitigation of targeted Sybil poisoning attacks.
We used the state-of-the-art FoolsGold algorithm (federated learning) as a starting point for SybilWall.

A simplified overview of the architecture of \algoname{} can be found in Figure \ref{fig:architecture}. During each training round, each node receives data about its (in)direct neighbors (step \circled{1}), which is stored in its local database (step \circled{2}, Section \ref{sec:local_database_updates}). This data, as well as the node's own model, is used by the aggregation function to produce the aggregated intermediary model (step \circled{3}, Section \ref{sec:aggregation_function}). This intermediary model is trained on the node's local private dataset to produce the trained model (step \circled{4}), which serves as the node's own model during the next training round.
Furthermore, for every neighbor, the node uses the probabilistic gossiping mechanism to select an entry from its local database (step \circled{5}, Section \ref{sec:propabilistic_gossip_mech}). Both this probabilistically selected model and the node's own model are used in the composition of the broadcast message (step \circled{6}, Section \ref{sec:message_composition}), which is transmitted to its neighbors.

\subsection{Aggregation function}
\label{sec:aggregation_function}
We improve upon the intuitive direction of FoolsGold (Section \ref{sec:foolsgold}), designed for inherently unscalable federated learning. By exploiting the high degree of similarity between Sybil models, FoolsGold detects and diminishes the impact of Sybils on the training process. Based on this promising heuristic (Assumption \ref{as:sybils_similar_updates}), we adopt a modified version of FoolsGold as \algoname{}'s aggregation function (step \circled{3}). 
By doing so, \algoname{} is able to directly mitigate the \dense{}. In such a case, the aggregation function has knowledge of at least two directly connected Sybils producing highly similar models. These models are subsequently excluded during the aggregation as they are detected by the cosine similarity function.
Furthermore, our aggregation function improves on FoolsGold in two dimensions.


Firstly, we modify FoolsGold to always trust the aggregator. Adversaries are not capable of compromising a node's training dataset and their training function by Assumption \ref{as:limited_to_dl_api}. Therefore, nodes can trust and exclude their own work from the similarity function during aggregation. The aggregator's model is reintroduced into the aggregation with the maximum weight, after all other models have been assigned a score.

Secondly, \algoname{} supports incorporating additional data of indirect neighbors in the similarity function. This increases the probability of comparing the data of at least two Sybils, which do not necessarily have attack edges to the same honest node. By Assumption \ref{as:sybils_similar_updates}, these Sybils will produce similar data each training round. Consequently, in the case where an honest node is only connected to a single attack edge, the incorporation of data from indirect neighbors might lead to the attack edge's mitigation.
Although data from indirect neighbors are included in the similarity function, only data from direct neighbors are considered for aggregation.
Note that obtaining data from indirect neighbors is not supported by the default decentralized learning API, but is facilitated through gossiping by the probabilistic gossiping mechanism.

\subsection{Probabilistic gossiping mechanism}
\label{sec:propabilistic_gossip_mech}
We devised a probabilistic gossiping mechanism (step \circled{5}), which allows data dissemination among indirect neighbors. By doing so, the sensitivity of the aggregation function improves as the amount of reference material for the similarity function increases. 
Furthermore, this enables \algoname{} to mitigate the \distributed{}, as the probabilistic gossiping mechanism provides the aggregation function with gossiped data from a wider scope of nodes.
The gossiped data consists of the model history $h^T_i$ of some node $i$ in round $T$, which is defined by $h^T_i = \sum_{t = 0}^T w_i^t$, where $w_i^t$ is a trained model produced by node $i$ in round $t$. 

First, let us define the method in which model histories are selected to be propagated to a neighboring node, for which \algoname{} employs a weighted random selection algorithm. 

More specifically, let $\mathcal{H}_i$ denote the local database of model histories of node $i$. $\mathcal{H}_i$ consists of a list of tuples, each in the form $\langle p, h, r, d, f \rangle \in \mathcal{H}_i$. Here, $h$ stands for the model history of node $p$ and $r$ signifies the identifier of the synchronous training round from which the model history originates. $d$ represents the distance that the model history has traveled, counted in the number of propagations. The term $f$ refers to the neighbor of node $i$ that provided this particular model history.
Given the current node $i$ and its neighboring node $j$, let the filtered database of model histories $\mathcal{H}_i^j$ be defined as $H_i^j = \{(p, h, r, d, f) \in H_i\ |\ p \notin \{i, j\} \wedge f \neq j \}$. 
This filtered database is used in a weighted random selection to determine which model history will be gossiped to node $j$.

To perform the weighted random selection, the entries of the filtered database of model histories are first assigned weights. These weights directly correspond to the traveled distance $d$ and are assigned according to the exponential distribution:
\begin{equation}
    P(d) = \lambda e^{-\lambda d}
\end{equation}
where $\lambda$ can be considered a hyperparameter representing the relevance of propagating the model history of distant nodes.
The choice for the exponential distribution is not arbitrary, as it prioritizes the propagation of the model history of nearby nodes over that of distant nodes. This approach assumes that the \sparse{} is mitigated through a natural dampening effect. This natural dampening effect originates from the repeated train-aggregate loop (Figure \ref{fig:trainaggregateloop}) on each node, causing the influence of a Sybil to fade as the distance to the attack edge increases. After the weights have been assigned to the filtered database of model histories, a weighted random selection is performed to select the model history that is propagated. 

\subsection{Local database updates}
\label{sec:local_database_updates}
A node's local database of model histories can be updated in two distinct methods (step \circled{2}). First, if a node $i$ receives a model history through gossiping from some other node $j$, which it has not seen before, it is added to $i$'s local database. Second, if node $i$ receives a model history from some node $k$ that is more recent than the prior model history of $k$ known to node $i$, it is updated accordingly. Note that the model histories of direct neighbors are updated every round, as each training round will result in a more recent model history. It is possible that a node's local database of model histories may grow to a significant size over time, resulting in a decrease in performance during aggregation. In such a scenario, \algoname{} can drop outdated model histories to prevent performance loss.

\subsection{Message composition} 
\label{sec:message_composition}
The previously described probabilistic gossiping mechanism requires model histories to be propagated to neighbors, which can be maliciously manipulated if implemented naively. As such, adversaries could exploit this by increasing the similarity of two targeted nodes, thereby potentially reducing the relative similarity among its Sybils. To mitigate this vulnerability, \algoname{} replaces the original model communication discussed in Section \ref{sec:dl} with a more secure communication scheme, which involves the use of signed histories.

To enable the use of signatures (secure by Assumption \ref{as:cryptographic_primitives_safe}), the model history and the corresponding signature are constructed on the originating node. By doing so, any node can propagate signed model histories of (in)direct neighbors. However, this induces additional communication overhead since the trained model, the signed model history, and a signed gossiped model history all need to be communicated to neighbors every training round. We decrease these communication costs by omitting the trained model, as it can be inferred from the comparison of the two most recent model histories.

More specifically, \algoname{} composes messages (step \circled{6}) such that a message $m_{i \to j}^T$ from node $i$ to $j$ in round $T$ can be decomposed into $\langle M_i^T, M_k^r, d_{k \to i} + 1 \rangle$. In this decomposition, $M_i^T$ represents the model history information originating from node $i$ in round $T$, $M_k^r$ is the gossiped model history from node $k$ originating from round $r$ and $d_{k \to i}$ is the distance model history $M_k^r$ has traveled so far. $M_i^T$ can be further decomposed into $\langle h_i^T, i, T, S_i(h_i^T, T) \rangle$, where $h_i^T$ is the model history of node $i$ in round $T$ and $S_i$ is node $i$'s signature function. Note that all nodes construct their own model histories through cumulative summation of the trained model.

\subsection{Downtime tolerance}
In contrast to a pull-based communication scheme, in which nodes could stochastically request a (distant) node's signed model history, \algoname{} supports arbitrary downtime or the presence of private networks, both resulting in unreachable nodes. 
Moreover, a pull-based communication scheme would enable trivial manipulation of model histories, allowing adversaries to make Sybils seem more diverse.
Using \algoname{}'s communication scheme, nodes are not responsible for the propagation of their own model history. Therefore, consistent reachability is not a necessity to proceed the training process.

In the event that a node experiences downtime, its aggregation function will start operating properly again once the node is online again and skips an additional training round. By doing so, it can obtain two subsequent model histories from its neighbors. This allows for inference of the trained model, since the difference between the two received model histories corresponds to the trained model required for aggregation.

\section{Evaluation}
\label{sec:evaluation}
We evaluate \algoname{} by answering the following questions: \textit{(1) How does the complexity of the dataset and the model affect the performance of \algoname{}? (2) How does \algoname{} perform compared to other existing algorithms? (3) How does the attack density $\phi$ influence the performance of \algoname{}? (4) What is the effect of the distribution of data among nodes on the performance of \algoname{}? (5) Can \algoname{} be further enhanced by combining it with different techniques?}

\begin{table}[t]
\centering
\caption{\label{tab:default_parameters}The default hyperparameters used during the evaluation of \algoname{}.}
\begin{tabular}{lr}
\textbf{Hyperparameter}                            & \textbf{Value}             \\ \hline
\multicolumn{1}{l}{\# honest nodes} & \multicolumn{1}{r}{99} \\ \rowcolor[HTML]{EFEFEF}
\multicolumn{1}{l}{Attack edge density $\phi$} & \multicolumn{1}{r}{1} \\
\multicolumn{1}{l}{Gossip mechanism parameter $\lambda$} & \multicolumn{1}{r}{0.8} \\ \rowcolor[HTML]{EFEFEF}
\multicolumn{1}{l}{Dirichlet concentration parameter $\alpha$} & \multicolumn{1}{r}{0.1} \\ 
\multicolumn{1}{l}{Max node degree $d$} & \multicolumn{1}{r}{8} \\ \rowcolor[HTML]{EFEFEF}
\multicolumn{1}{l}{Local epochs} & \multicolumn{1}{r}{10} \\
\multicolumn{1}{l}{Batch size} & \multicolumn{1}{r}{8} \\ \hline
\end{tabular}
\end{table}

\subsection{Experimental setup}
We implemented \algoname{} in Python3 in the context of a fully operational decentralized learning system for experimental evaluation and is online available \cite{code}. We have used the PyTorch \cite{pytorch} library for the training of machine learning models. Regarding communication between individual nodes, we leveraged IPv8 \cite{pyipv8}, which provides an API for constructing network overlays in order to simulate P2P networks. Furthermore, we adopted the Gumby library \cite{gumby} as the experimental execution framework, which was specifically designed for sophisticated experiments with IPv8 involving many nodes. All experiments were performed on the Distributed ASCI Supercomputer 6 (DAS-6) \cite{das6}. Each node in the compute cluster has access to a dual 16-core CPU, 128 GB RAM, and either an A4000 or A5000 GPU. Furthermore, all default hyperparameters for the experiments can be found in Table \ref{tab:default_parameters}. Except where mentioned otherwise, these default hyperparameters define the configuration of all experiments.

In all experiments, we measure the accuracy by averaging the accuracy of the models of all honest nodes. Simultaneously, we measure the success rate of the attacker by averaging the attack score achieved on the models of all honest nodes. The attack score is defined as the accuracy that a model obtains on the altered segment of the data obtained by transforming the test dataset by the data transformation functions defined in Equation \ref{eq:label_flip} and \ref{eq:backdoor}. Note that both metrics are measured each round directly after aggregation.

\begin{table}[t]
\centering
\caption{\label{tab:datasets}The datasets used in the evaluation of \algoname{}.}
\begin{tabular}{lll}
\textbf{Dataset}                            & \textbf{Model}                                      & \textbf{Learning rate}         \\ \hline
\multicolumn{1}{l}{MNIST\cite{mnist}}        & \multicolumn{1}{l}{Single soft-max layer} & \multicolumn{1}{l}{$\eta = 0.01$ \cite{foolsgold}} \\ \rowcolor[HTML]{EFEFEF}
\multicolumn{1}{l}{FashionMNIST \cite{fashionmnist}} & \multicolumn{1}{l}{Single soft-max layer} & \multicolumn{1}{l}{$\eta = 0.01$ \cite{foolsgold}} \\ 
\multicolumn{1}{l}{CIFAR-10 \cite{cifar10}}     & \multicolumn{1}{l}{LeNet-5 \cite{lenet5}}               & \multicolumn{1}{l}{$\eta = 0.004$ \cite{cifarlearningrate}} \\ \rowcolor[HTML]{EFEFEF}
\multicolumn{1}{l}{SVHN \cite{svhn}}         & \multicolumn{1}{l}{LeNet-5 \cite{lenet5}}               & \multicolumn{1}{l}{$\eta = 0.004$ \cite{cifarlearningrate}} \\ \hline
\end{tabular}
\end{table}

\subsubsection{Datasets}
The datasets used during evaluation can be found in Table \ref{tab:datasets}. These datasets were chosen for a number of reasons. First of all, MNIST \cite{mnist} is a widely used dataset for the evaluation of machine learning algorithms \cite{onlymnist, onlymnist2, bristle}, serving as an adequate baseline algorithm for \algoname{}. FashionMNIST was developed as a more challenging variant of MNIST, thus serving as an ideal candidate to demonstrate the direct correlation between the complexity of classification tasks and the performance of \algoname{}. The choice for SVHN and CIFAR-10 is motivated by the increased complexity of the models required to obtain satisfactory accuracy, which may affect the performance of \algoname{}. The use of complex multilayer models in evaluation is frequently overlooked in related work or is performed only on a single dataset \cite{bristle, onlymnist, onlymnist2, onlycifar, onlycifar2}. Moreover, when multilayer models are used, they are regularly pre-trained and trained solely through transfer learning \cite{transferlearning, bristle}. While we recognize that all the datasets employed in this experimental evaluation focus on image classification, we argue that focusing on image classification is justifiable as it is known as a well-established task in machine learning. Furthermore, image classification frequently serves as a benchmark for evaluating distributed machine learning algorithms \cite{modest, bristle, onlymnist, onlymnist2, onlycifar, onlycifar2}, and there exists a variety of widely available datasets constructed specifically for this task. 

The models that are trained using the aforementioned datasets can also be found in Table \ref{tab:datasets}, as well as the corresponding learning rate $\eta$. Note that all the models in this evaluation are neural networks and are trained using stochastic gradient descent (SGD).

\subsubsection{Data distribution}

\begin{figure}
    \centering
    \includegraphics[width=0.7\linewidth]{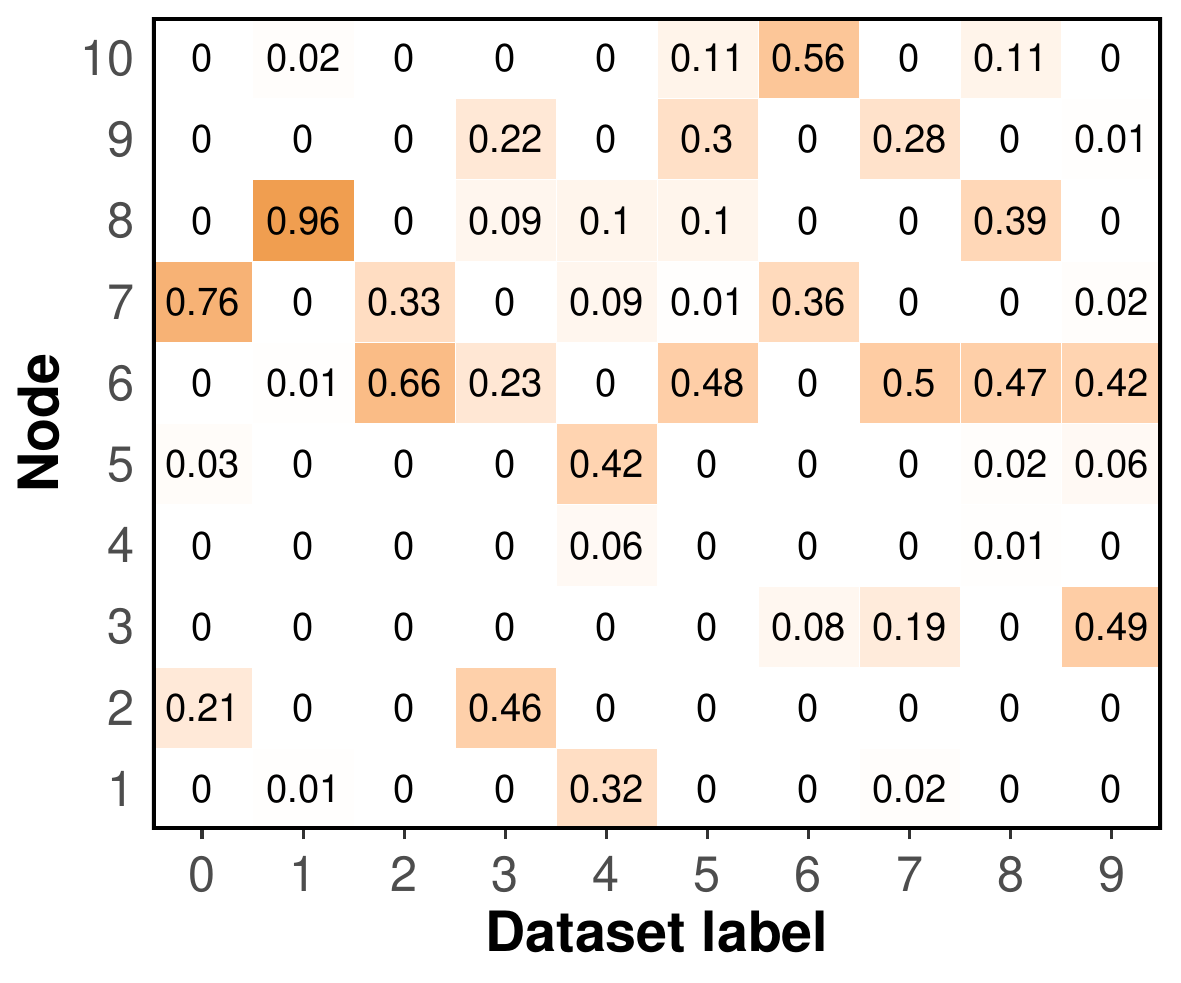}
    \caption{Example distribution for non-i.i.d. data generated with the Dirichlet distribution with concentration parameter $\alpha = 0.1$ for 10 nodes and a dataset containing 10 labels.}
    \label{fig:example_distribution}
\end{figure}

The aforementioned datasets are designed for centralized machine learning and require to be distributed among the participating nodes. During our evaluations, we assume that the data is \textit{ not} identically and independently distributed (non-i.i.d.), which more closely resembles real-world data than uniformly distributed data (i.i.d) \cite{9094657, pmlr-v119-hsieh20a}. Some studies employ the use of a K-shard data distribution \cite{bristle, shardandclassdatadistribution, sharddistrib, federatedlearning} or simply assign each node a predefined number of classes of the training data \cite{shardandclassdatadistribution, datadistribpaper}. However, we utilize the Dirichlet distribution \cite{dirichlet}, which has recently gained more popularity for generating non-i.i.d. data distributions \cite{modest, dirichletgood, dirichletgood2}. More specifically, given the \textit{concentration parameter} $\alpha$, we compute for each class the fraction of data every node possesses using the Dirichlet distribution, creating seemingly naturally unfair and irregular data distributions. Lower values of $\alpha$ result in more non-i.i.d. data. Figure \ref{fig:example_distribution} illustrates an example distribution for a dataset of 10 labels distributed over 10 nodes with a concentration parameter of $\alpha = 0.1$.

\begin{figure}
    \centering
  \subfloat[\label{fig:datasets-labelflip-accuracy}Accuracy \labelflip{}]{%
       \includegraphics[width=0.5\linewidth]{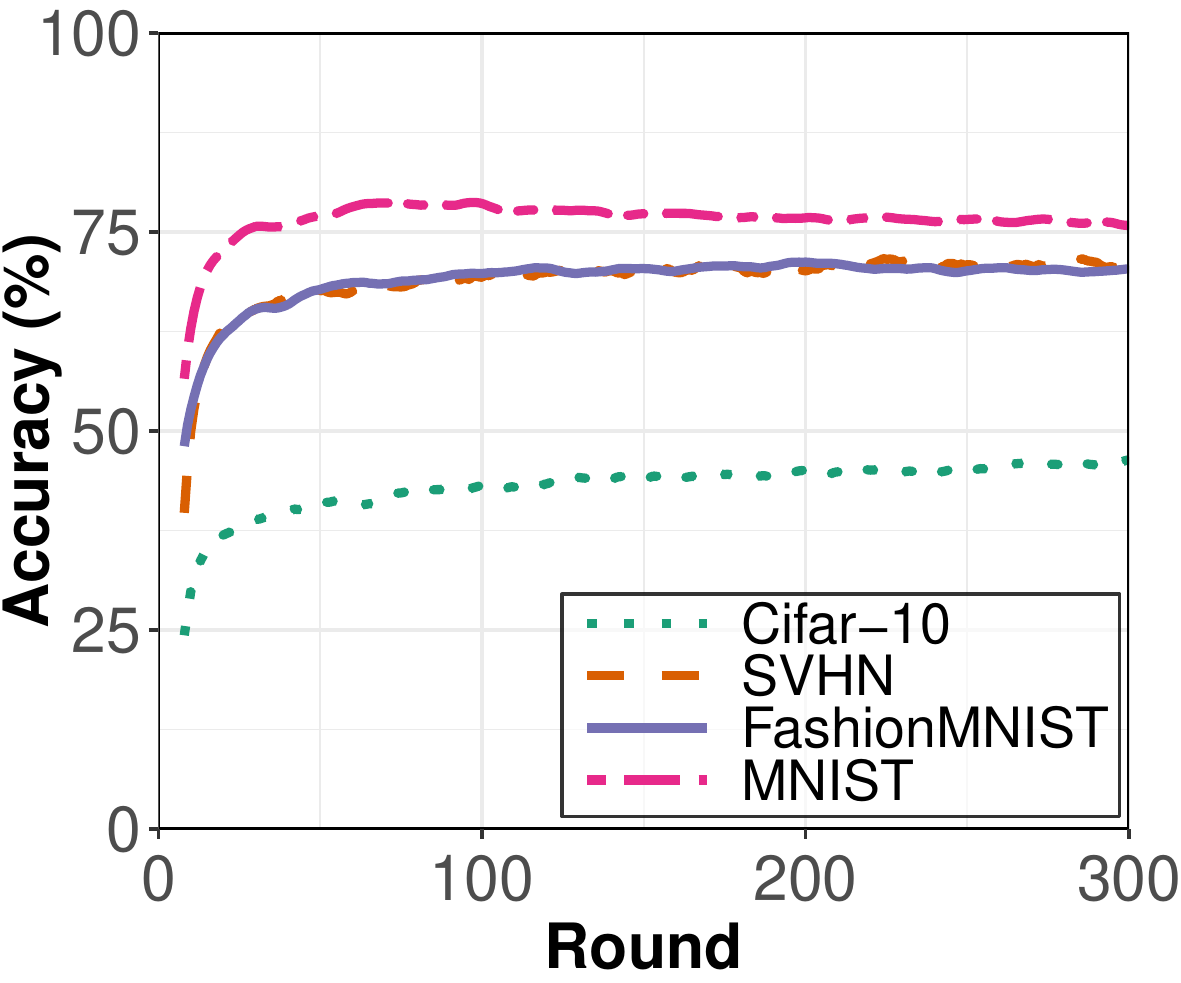}}
  \subfloat[\label{fig:datasets-labelflip-attack-score}Attack score \labelflip{}]{%
        \includegraphics[width=0.5\linewidth]{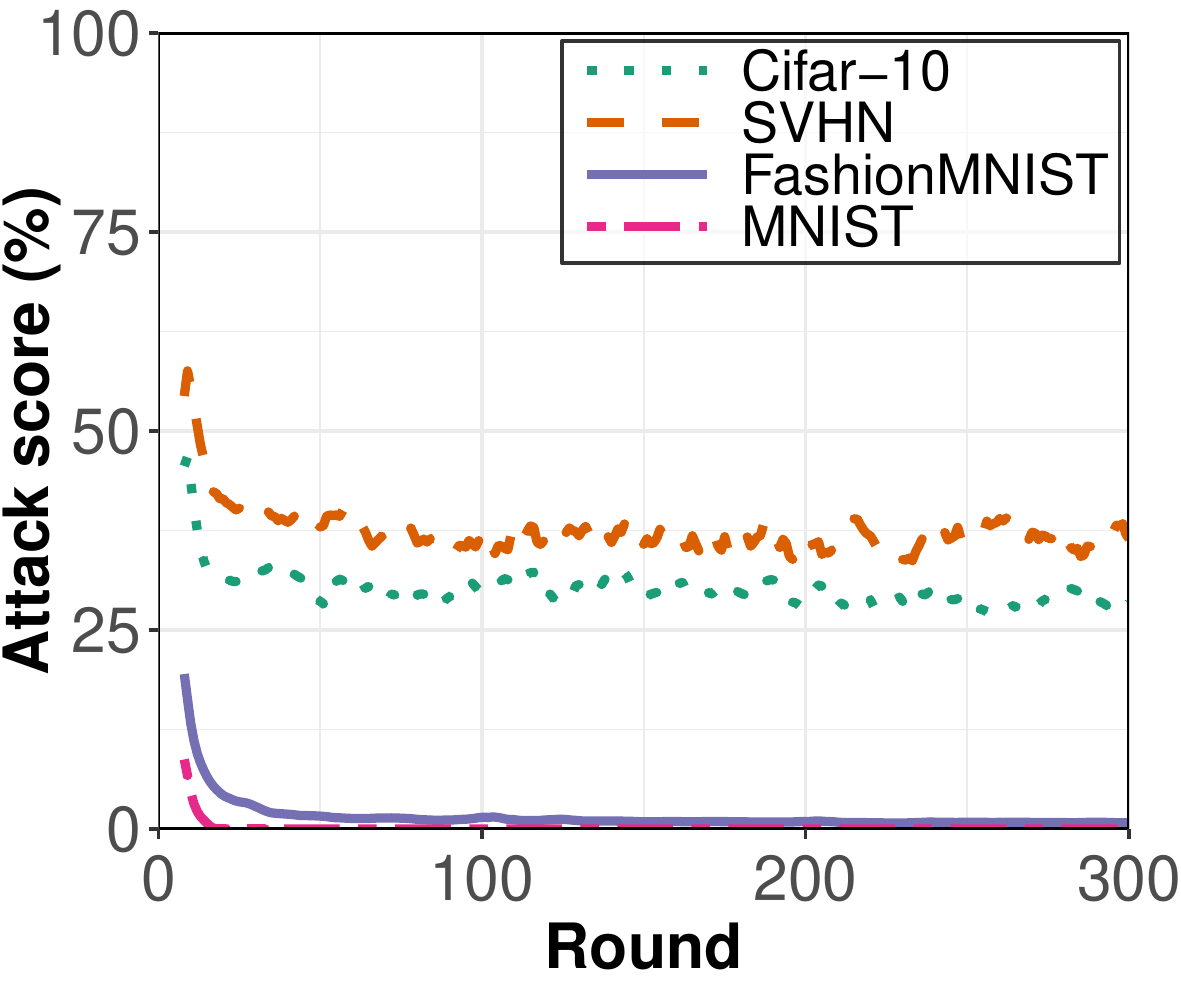}}
    \hfill
  \subfloat[\label{fig:datasets-backdoor-acc}Accuracy \backdoor{}]{%
        \includegraphics[width=0.5\linewidth]{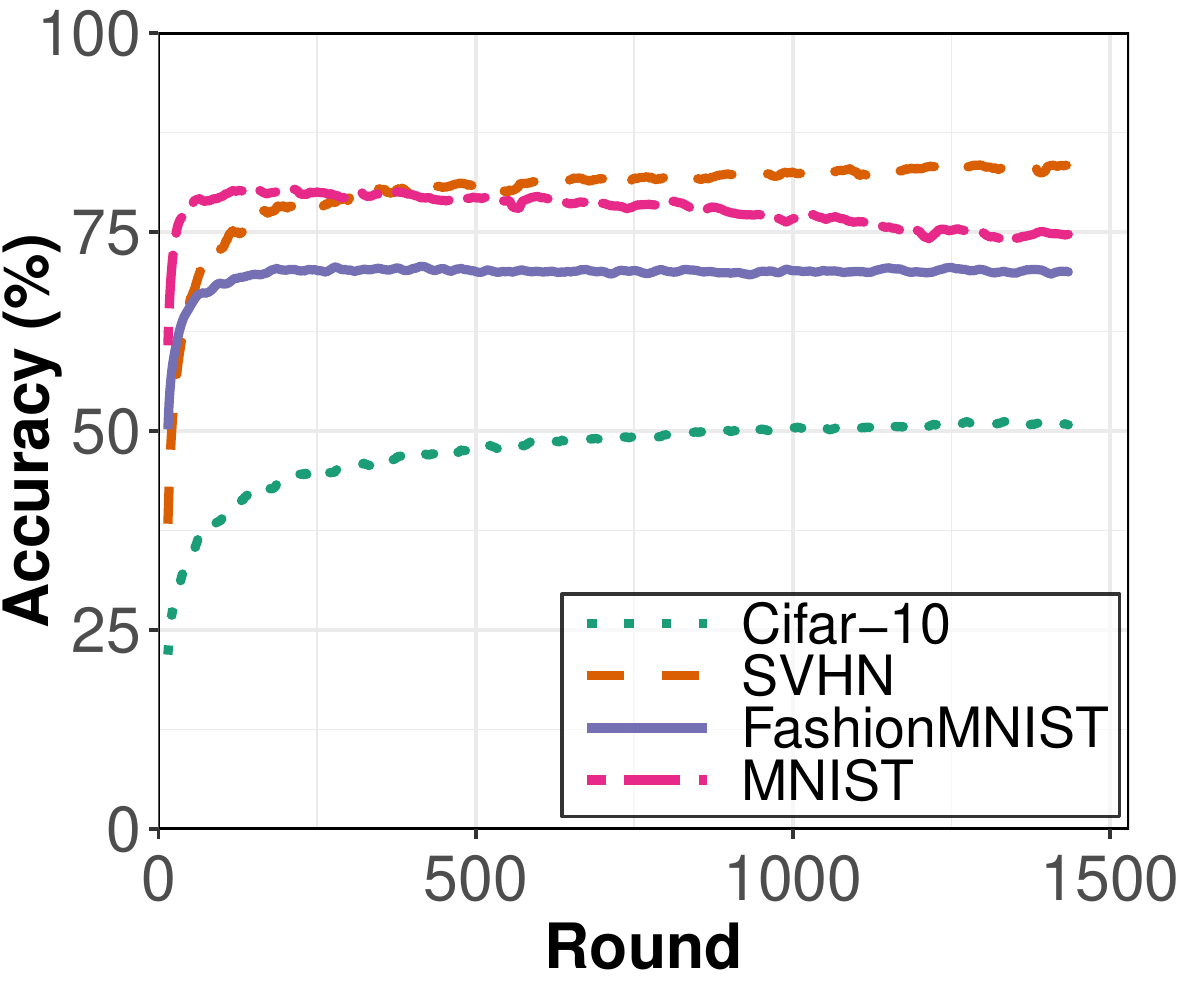}}
    \hfill
  \subfloat[\label{fig:datasets-backdoor-att}Attack score \backdoor{}]{%
        \includegraphics[width=0.5\linewidth]{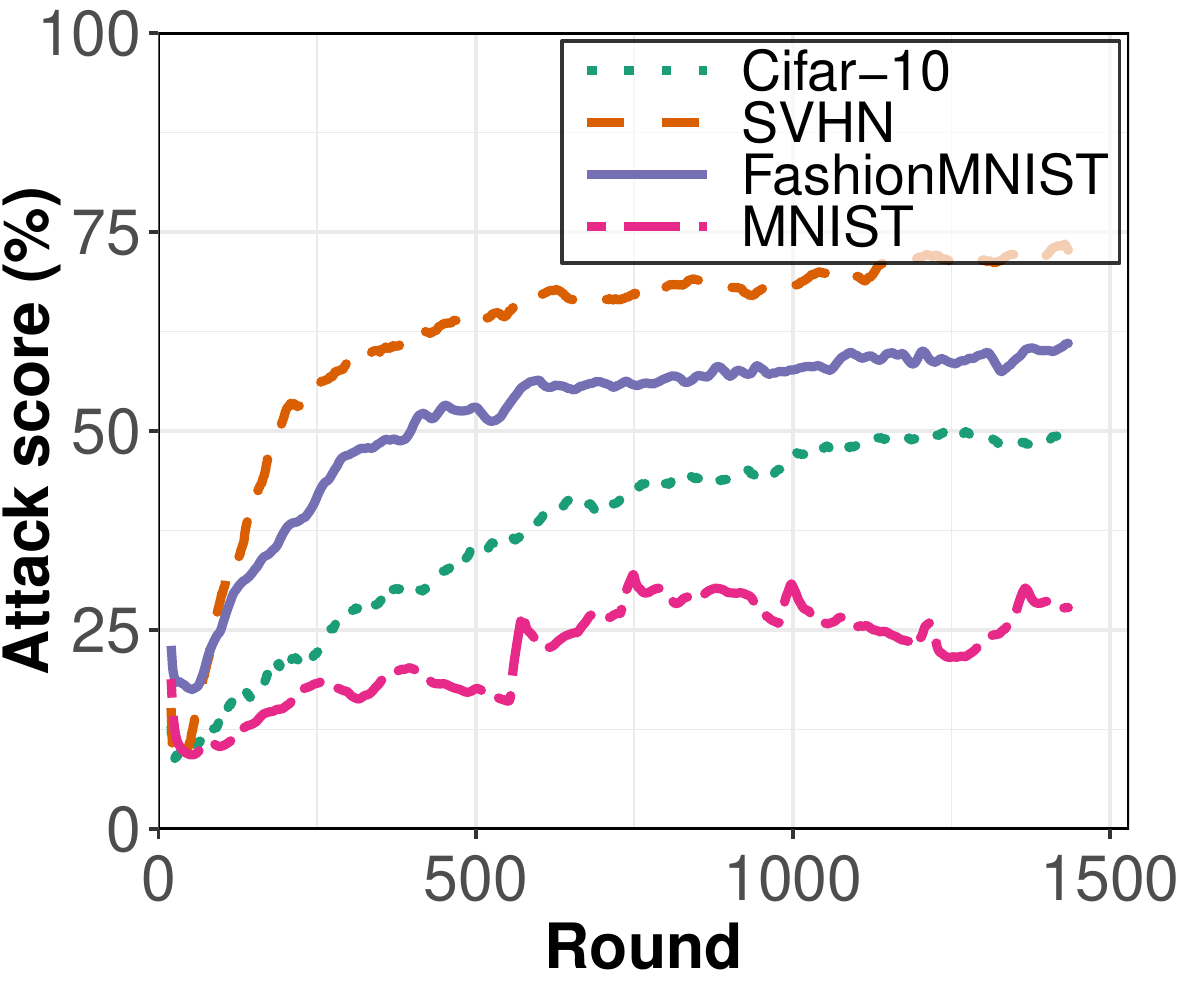}}
  \caption{\label{fig:datasets}Accuracy and attack score for the \labelflip{} (300 rounds) and the \backdoor{} (1450 rounds) on different datasets.}
\end{figure}

\begin{figure*}[t]
    \centering
    \subfloat[Accuracy \labelflip{}]{
        \includegraphics[width=0.245\linewidth]{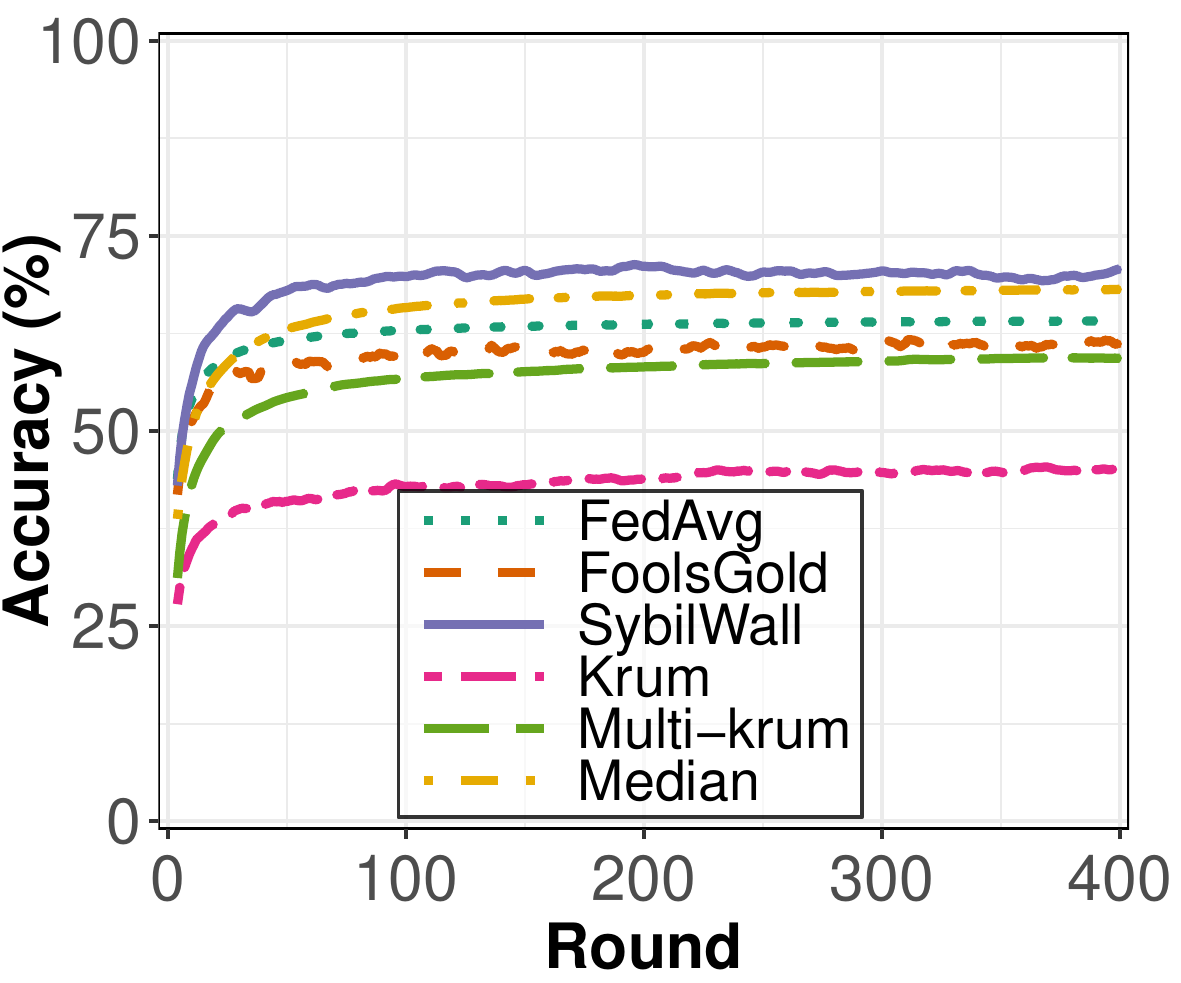}}
  \subfloat[Attack score \labelflip{}]{
        \includegraphics[width=0.245\linewidth]{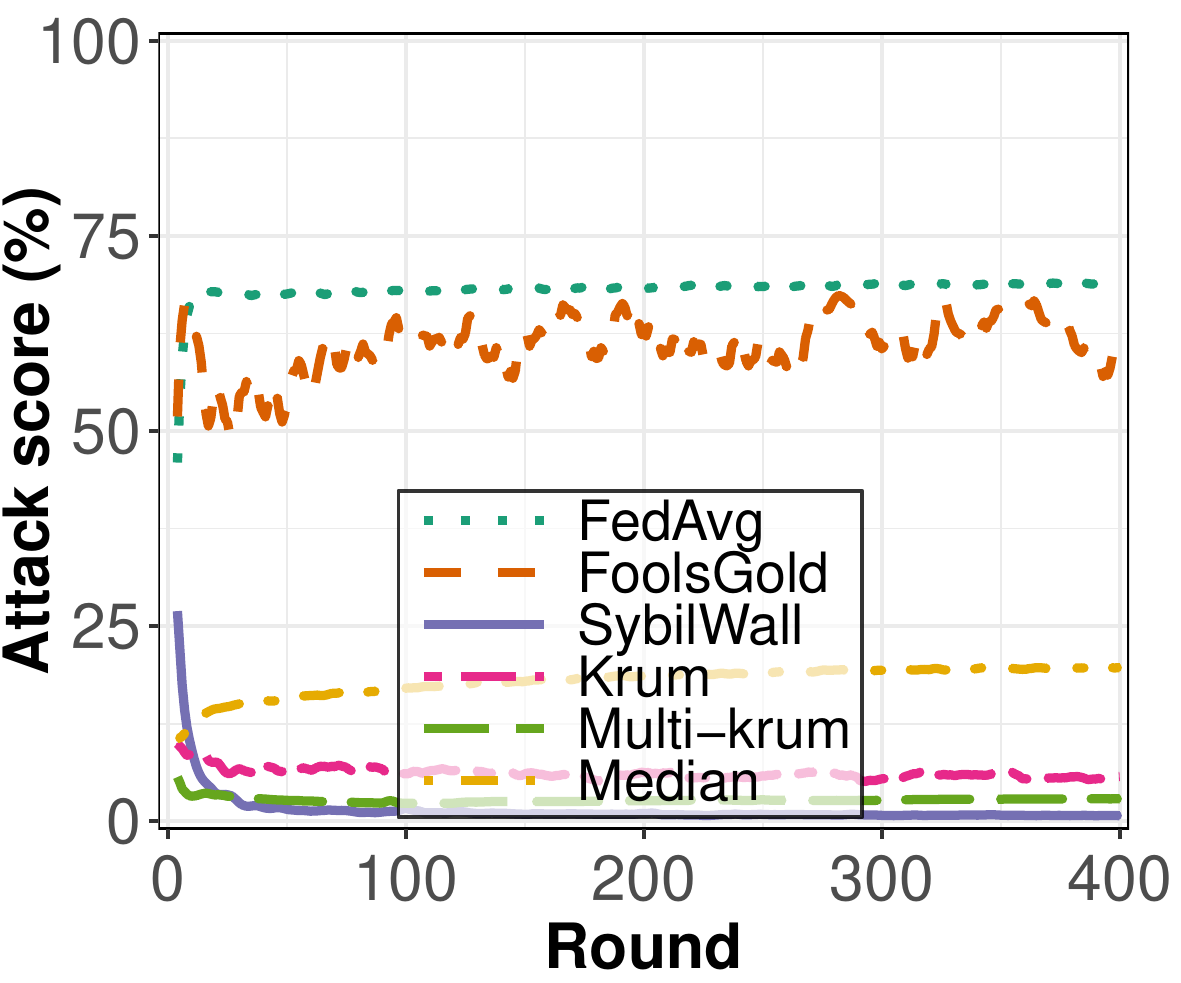}}
  \subfloat[Accuracy \backdoor{}]{%
        \includegraphics[width=0.245\linewidth]{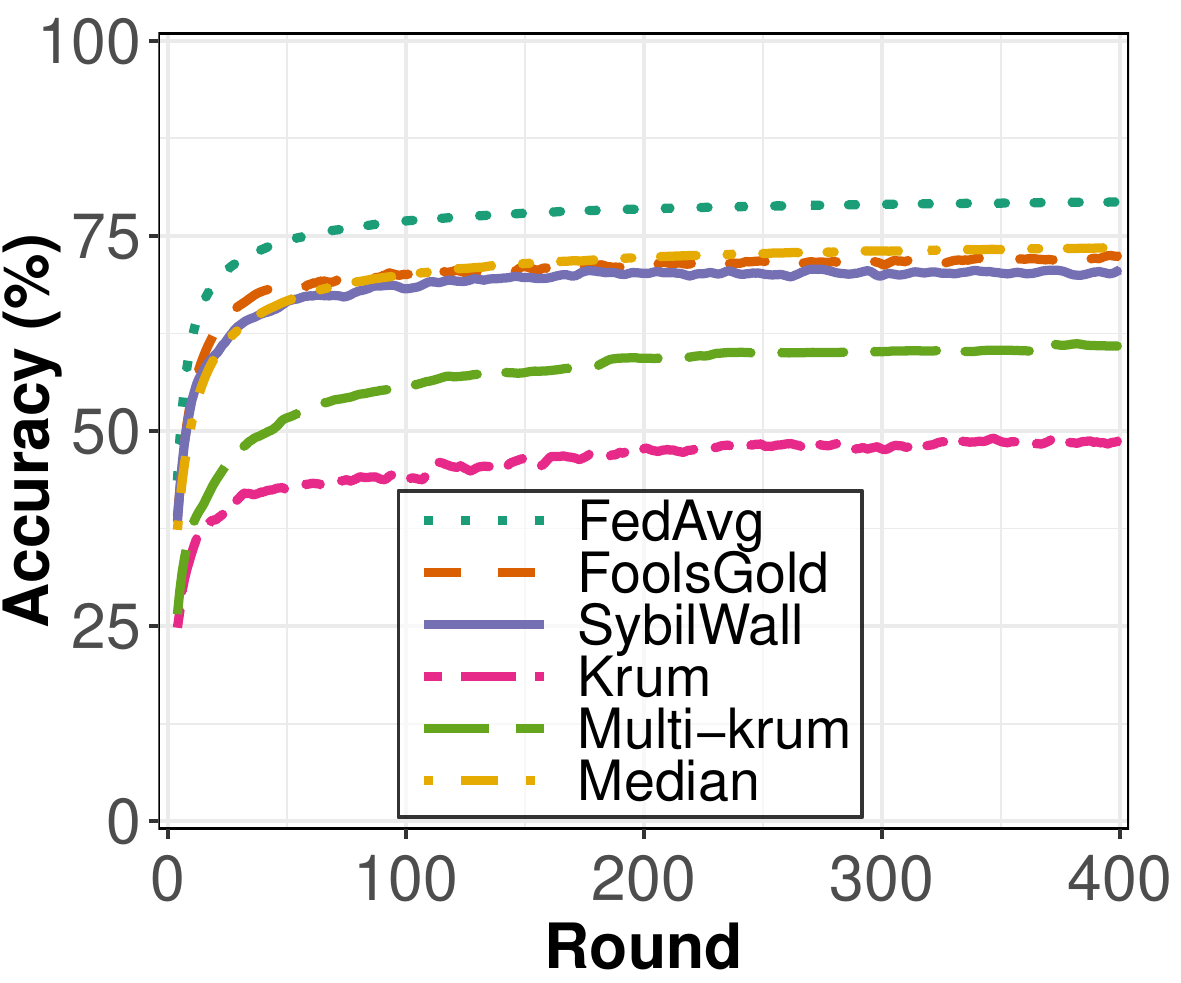}}
  \subfloat[Attack score \backdoor{}]{
        \includegraphics[width=0.245\linewidth]{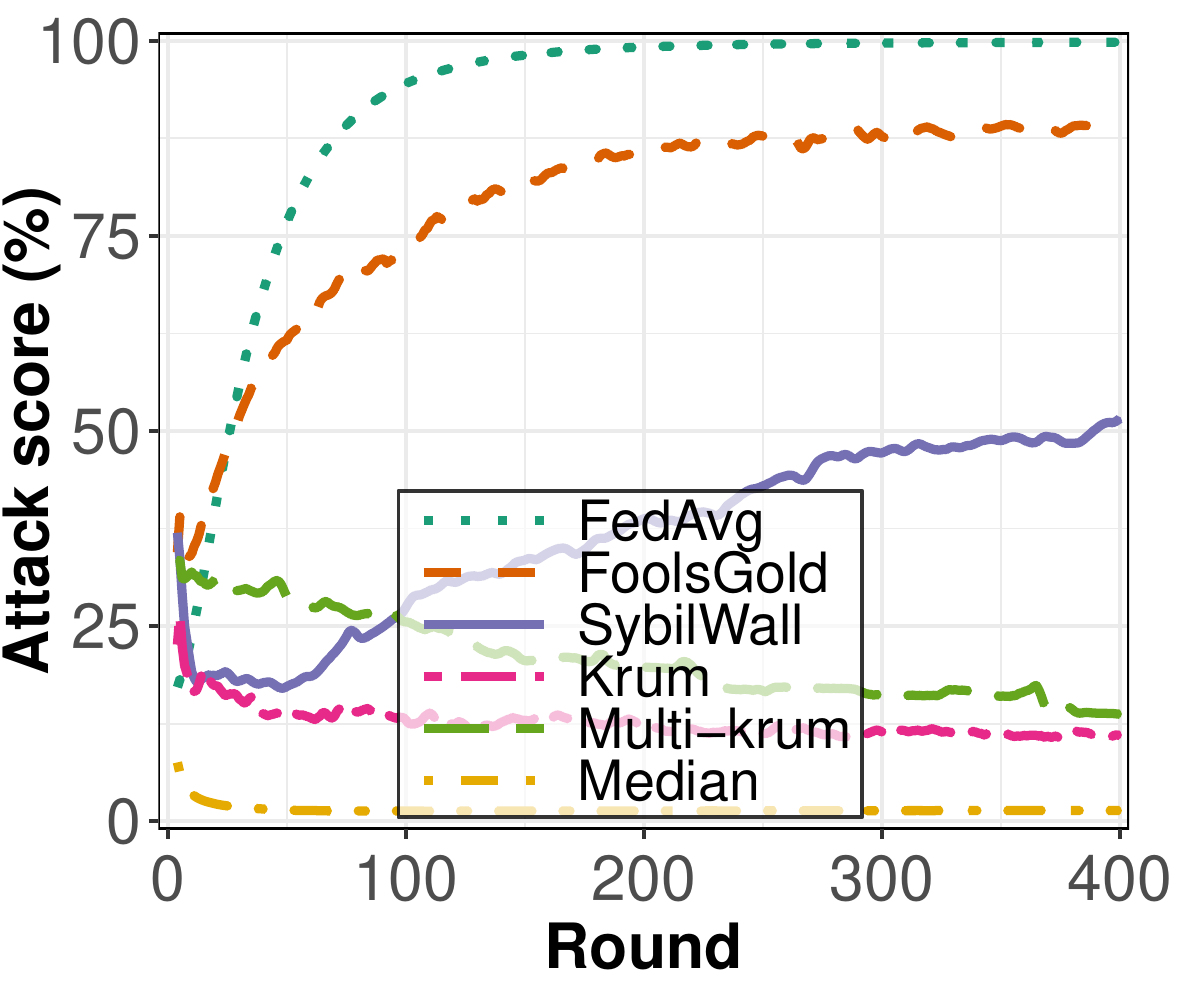}}
    \caption{\label{fig:techniques-phi1}Comparison of \algoname{} against different techniques on $\phi = 1$. Results generated using the FashionMNIST \cite{fashionmnist} dataset.}

    \subfloat[Accuracy \labelflip{}]{
        \includegraphics[width=0.245\linewidth]{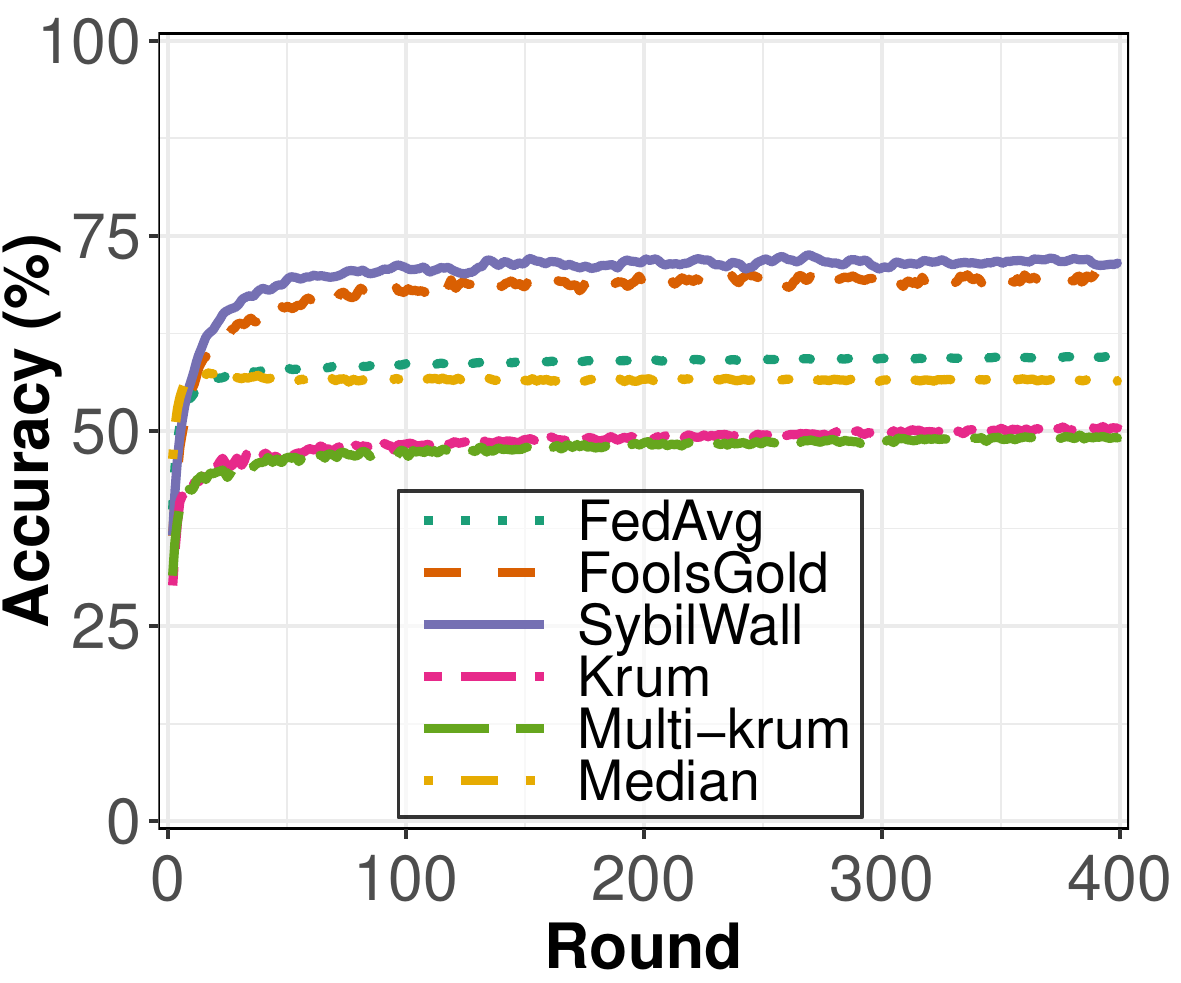}}
  \subfloat[Attack score \labelflip{}]{
        \includegraphics[width=0.245\linewidth]{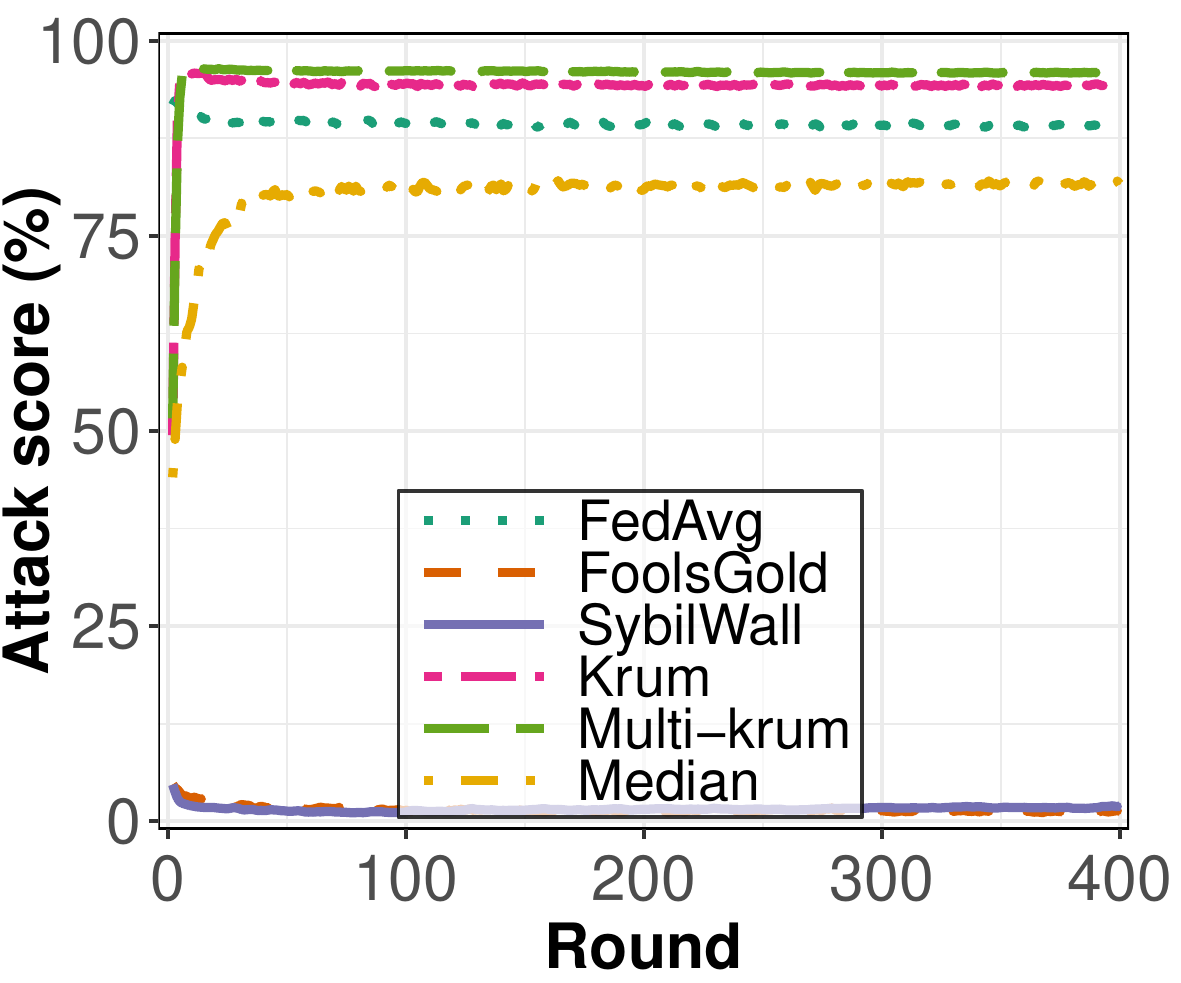}}
  \subfloat[Accuracy \backdoor{}]{%
        \includegraphics[width=0.245\linewidth]{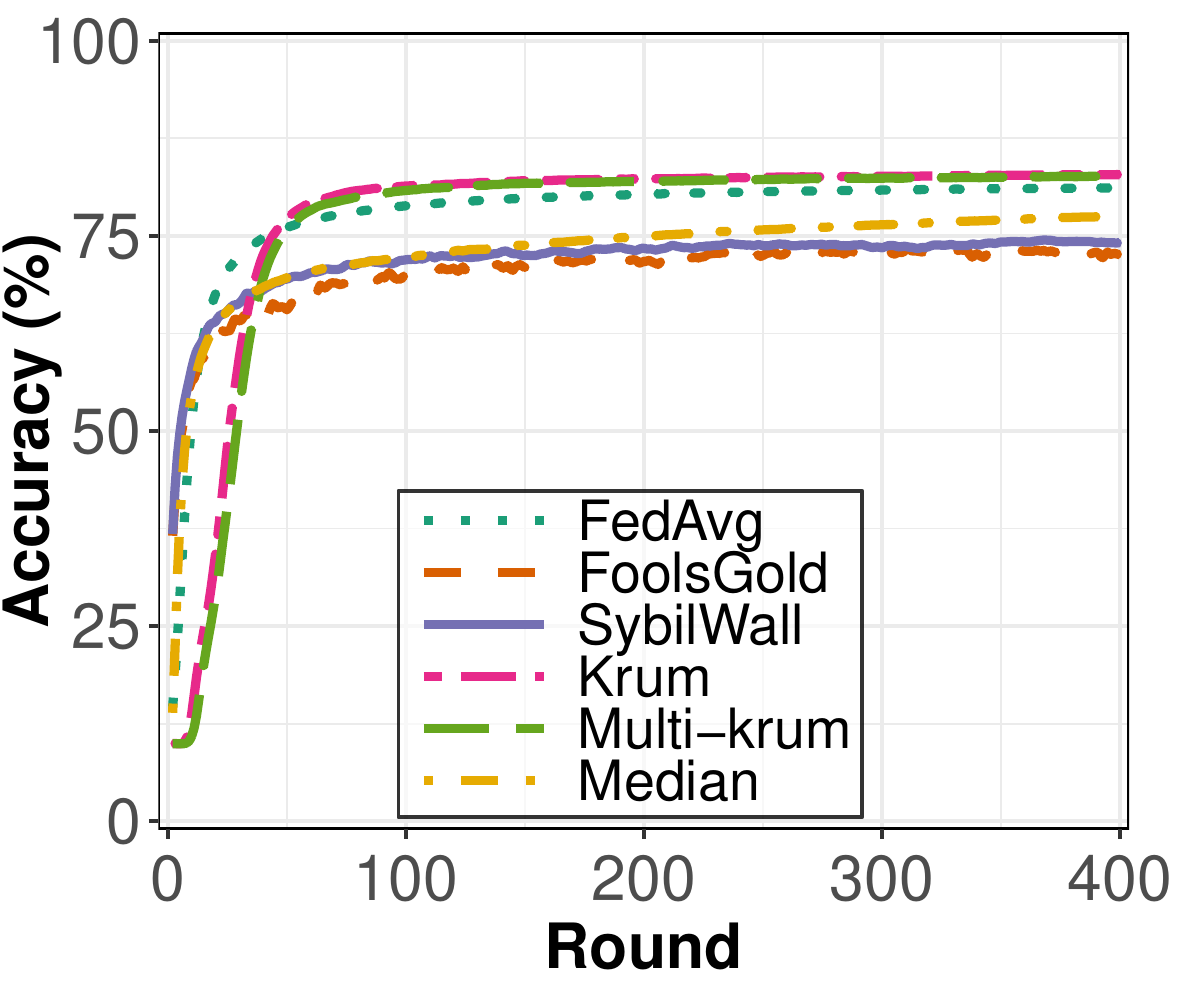}}
  \subfloat[Attack score \backdoor{}]{
        \includegraphics[width=0.245\linewidth]{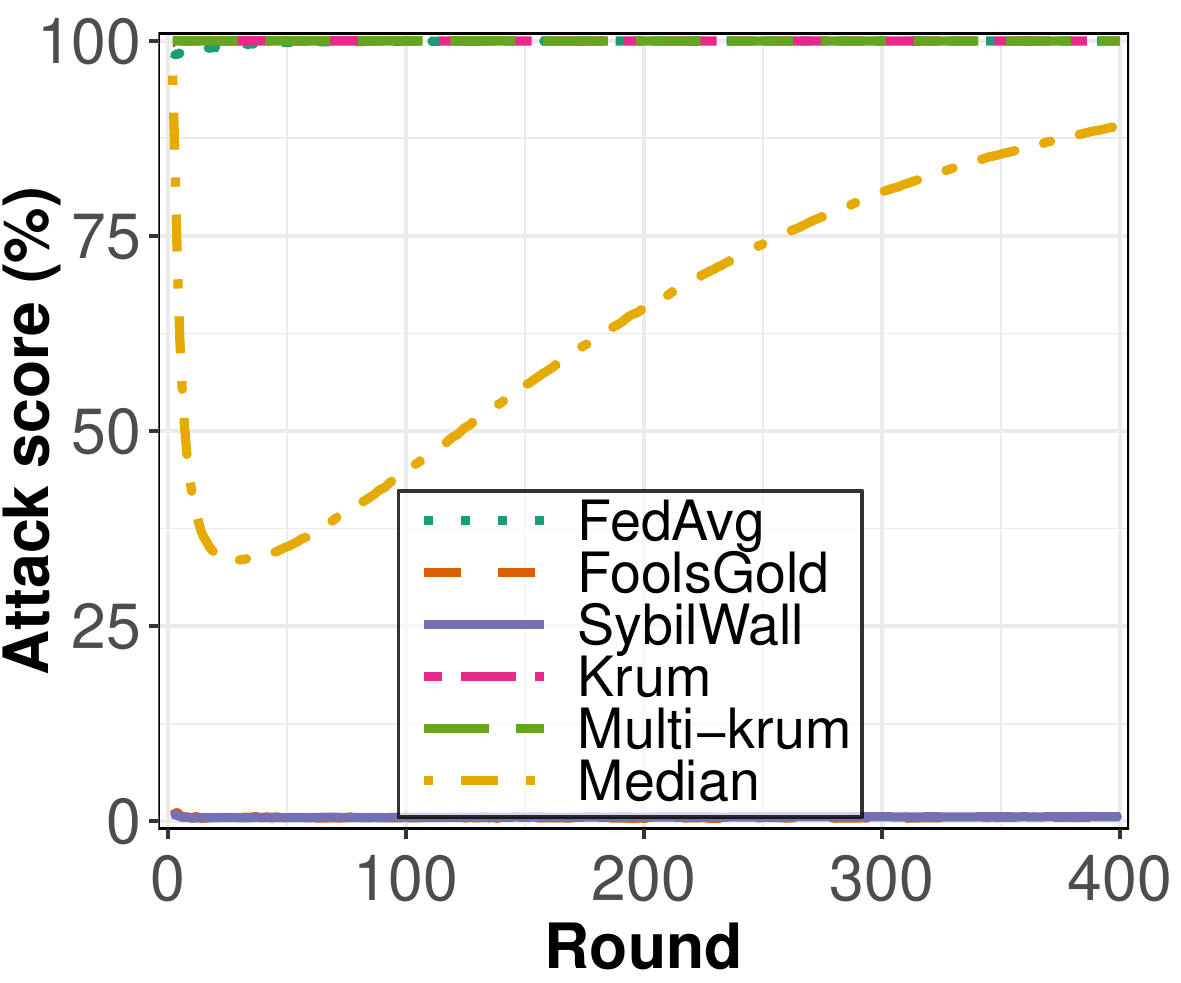}}
    \caption{\label{fig:techniques-phi4}Comparison of \algoname{} against different techniques on $\phi = 4$. Results generated using the FashionMNIST \cite{fashionmnist} dataset.}
\end{figure*}

\subsubsection{Network topology}
To generate the necessary network topologies, defining the relations between nodes, we employed \textit{random geometric graphs}. Random geometric graphs are constructed by randomly placing points, which correspond to nodes, on a grid. Two nodes are connected by an edge when the Euclidean distance between the corresponding points of these nodes is smaller than some predefined constant. To enforce the upper bound on a node's degree (Assumption \ref{as:degree_bounded}), random edges are removed from the random geometric graph, such that all nodes remain connected through a single connected component. The code used to generate these network topologies can be found in our published code repository \cite{code}. Furthermore, during our experiments, we assume a static network topology. That is, no nodes will leave or join the network during training, including Sybils. Lastly, we employ the SSP attack (Section \ref{sec:adverserial_attack_strategy}) as the adversarial strategy in the simulated Sybil attacks, since we hypothesize that more distant attack edges will result in a lower detection rate, thereby approximating the optimal attack scenario.

\subsection{Effect of dataset}
\label{sec:eval_dataset}
\subsubsection{Setup} We evaluated the performance of \algoname{} on different datasets, allowing us to observe how \algoname{} is affected by varying the complexity in both the dataset and the model. This experiment was carried out using the default parameters listed in Table \ref{tab:default_parameters} and using the datasets, models and learning rates listed in Table \ref{tab:datasets}.

\subsubsection{Results}
Figure \ref{fig:datasets} demonstrates the effect of varying the dataset on the trend of accuracy and attack score. We clearly observe that CIFAR-10, arguably the most challenging dataset used in this work, obtains a significantly lower accuracy compared to simpler datasets (Figure \ref{fig:datasets-labelflip-accuracy}), such as MNIST. This can be explained by difference in the complexity of the training samples, i.e. MNIST consists of grayscale images, while CIFAR-10 has RGB images. Moreover, samples in the CIFAR-10 dataset have more variety within a class, such as different backgrounds or different races of dogs.

A noteworthy observation with regard to the attack score of the \labelflip{} in Figure \ref{fig:datasets-labelflip-attack-score} is that datasets that require more sophisticated models, such as LeNet-5, are generally more susceptible to the \labelflip{} compared to simpler models, such as a single-layer neural network. 
The results suggest that the smaller number of trainable weights in the single-layer neural network cause the Sybil model histories to exhibit greater similarity relative to more sophisticated models when comparing model histories from different training rounds. 
Moreover, the greatly increased number of weights of the more sophisticated models allows for more diversity in the sum of the trained models.

Taking into account the results of the \backdoor{} depicted in Figures \ref{fig:datasets-backdoor-acc} and \ref{fig:datasets-backdoor-att}, it is apparent that all attack scores demonstrate an increasing trend over a prolonged period of time.
However, if we consider the convergence rate of both the accuracy and attack score, it becomes apparent that the attack score requires a substantially longer time period to reach convergence on most datasets.

\subsection{Comparison with different techniques}
\label{sec:eval_techniques}

\subsubsection{Setup}
We evaluate the performance of \algoname{} relative to a number of different techniques focused on mitigating Sybil poisoning attacks or Byzantine attacks in general. These techniques are the following:

\begin{enumerate}[i.]
    \item FedAvg \cite{federatedlearning}: naively averages all models. This algorithm was the first proposed federated learning aggregation algorithm and will serve as a baseline during our evaluation.
    \item FoolsGold \cite{foolsgold}: detects Sybils among its neighbors by assuming that Sybils produce highly similar models.
    \item Krum \cite{krum}: excludes Byzantine models by filtering for the model which has the smallest sum of Euclidean distances to its $n-f-2$ closest neighbors.
    \item Multi-krum \cite{krum}: similar to krum. Averages the $m$ models with the lowest sum of Euclidian distances to its $n-f-2$ closest neighbors.
    \item Median \cite{median}: computes the element-wise median of all models and thereby excludes outliers.
\end{enumerate}
During this experiment, we alternated the attack edge density $\phi \in \{1, 4\}$ and fixed the dataset on FashionMNIST.

\subsubsection{Results}
Figure \ref{fig:techniques-phi1} shows the results of \algoname{} compared to different techniques using attack edge density $\phi = 1$. We observe that \algoname{} always scores among the best performing algorithms in terms of accuracy. Especially considering the \labelflip{}, \algoname{} achieves the highest accuracy among all evaluated techniques. Furthermore, the results demonstrate that \algoname{} successfully mitigates the \labelflip{}, similarly to some of the other techniques evaluated. On the \backdoor{}, we observe that \algoname{} exhibits the same increasing trend as in the prior experiment on the effect of the datasets in Section \ref{sec:eval_dataset}; the initially low attack score gradually increases as training progresses. 

Figure \ref{fig:techniques-phi4} shows the results of \algoname{} compared to different techniques using a higher attack edge density $\phi = 4$. These results clearly demonstrate how most aggregation algorithms succumb under the use of a large-scale Sybil attack. Taking into account the accuracy of both \labelflip{} and \backdoor{}, we observe that the converged accuracy of most algorithms increases significantly when employing the \backdoor{}. This phenomenon can be explained by the fact that the adversary is not actively attempting to decrease the accuracy of the model. In fact, the adversary only attempts to insert an activation pattern, which was highly successful for the algorithms demonstrating an increased converged accuracy compared to the \labelflip{}. On the other hand, both FoolsGold and \algoname{} seem to be unaffected by both attacks. Regarding \algoname{}, this is likely caused by the integration of a modified version of FoolsGold, which was designed to mitigate the \dense{}.

Considering both the results in Figure \ref{fig:techniques-phi1} and \ref{fig:techniques-phi4}, Krum and Multi-krum algorithms surprisingly obtain a higher accuracy under a \backdoor{} with higher attack edge density compared to a lower attack edge density, while the accuracy of the other algorithms remain relatively constant in both scenarios. Lastly, we find that \algoname{} does not outperform all the alternative evaluated algorithms in all scenarios, but it is the only algorithm to consistently score among the best.

\subsection{Effect of attack edge density}
\label{sec:eval_attack_edge_density}

\subsubsection{Setup}
We evaluate \algoname{} in a number of different attack edge density configurations. This experiment aims to demonstrate the effect that an attacker can exercise on the network by creating different numbers of Sybils. MNIST is fixed as the dataset during this experiment and the attack edge density $\phi$ is varied within the range $\phi \in [0.1, 2]$.

\subsubsection{Results}
\begin{figure}
    \centering
  \subfloat[\label{fig:attack-densities-labelflip-accuracy}Accuracy \labelflip{}]{%
       \includegraphics[width=0.5\linewidth]{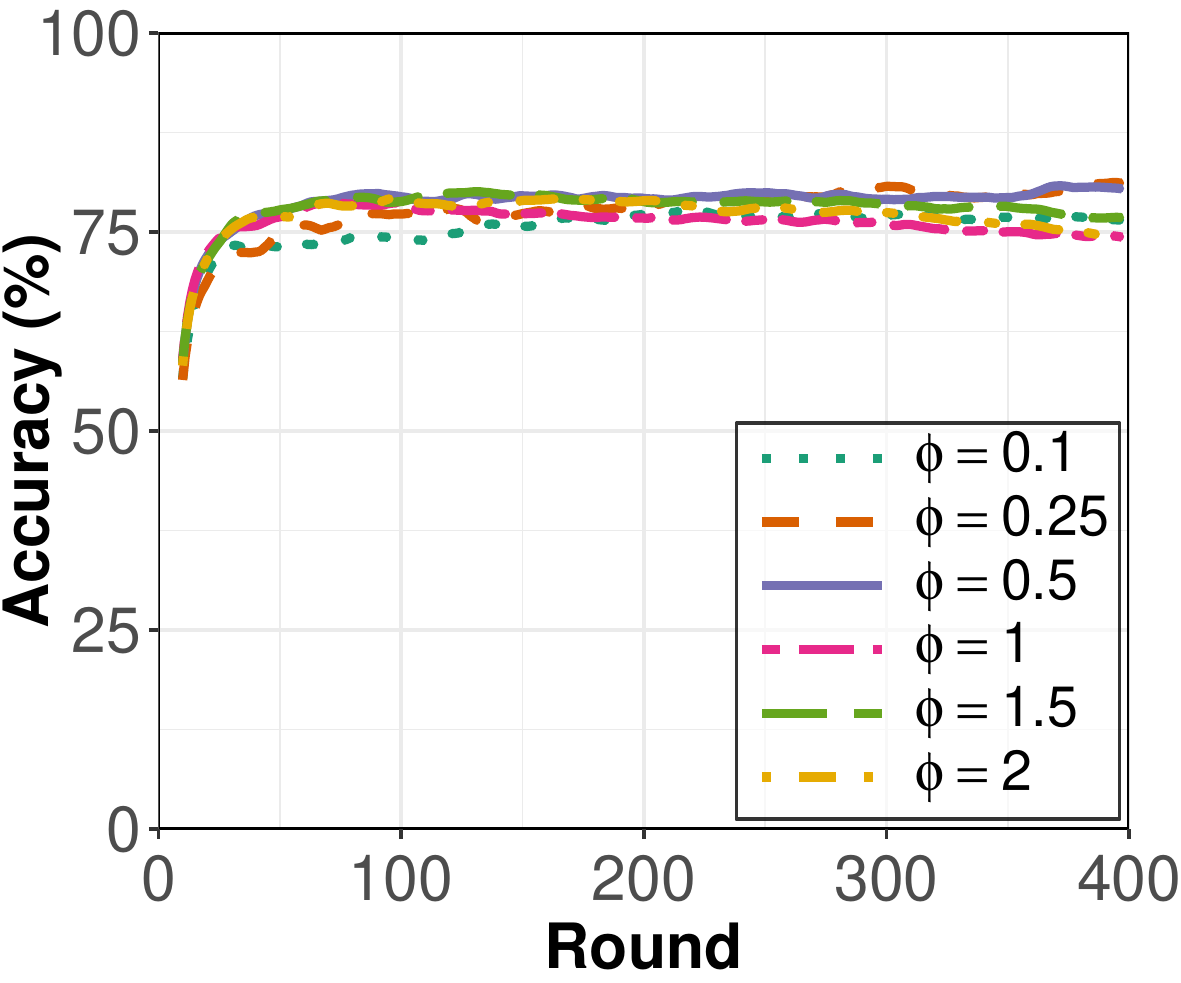}}
  \subfloat[\label{fig:attack-densities-labelflip-attack}Attack score \labelflip{}]{%
        \includegraphics[width=0.5\linewidth]{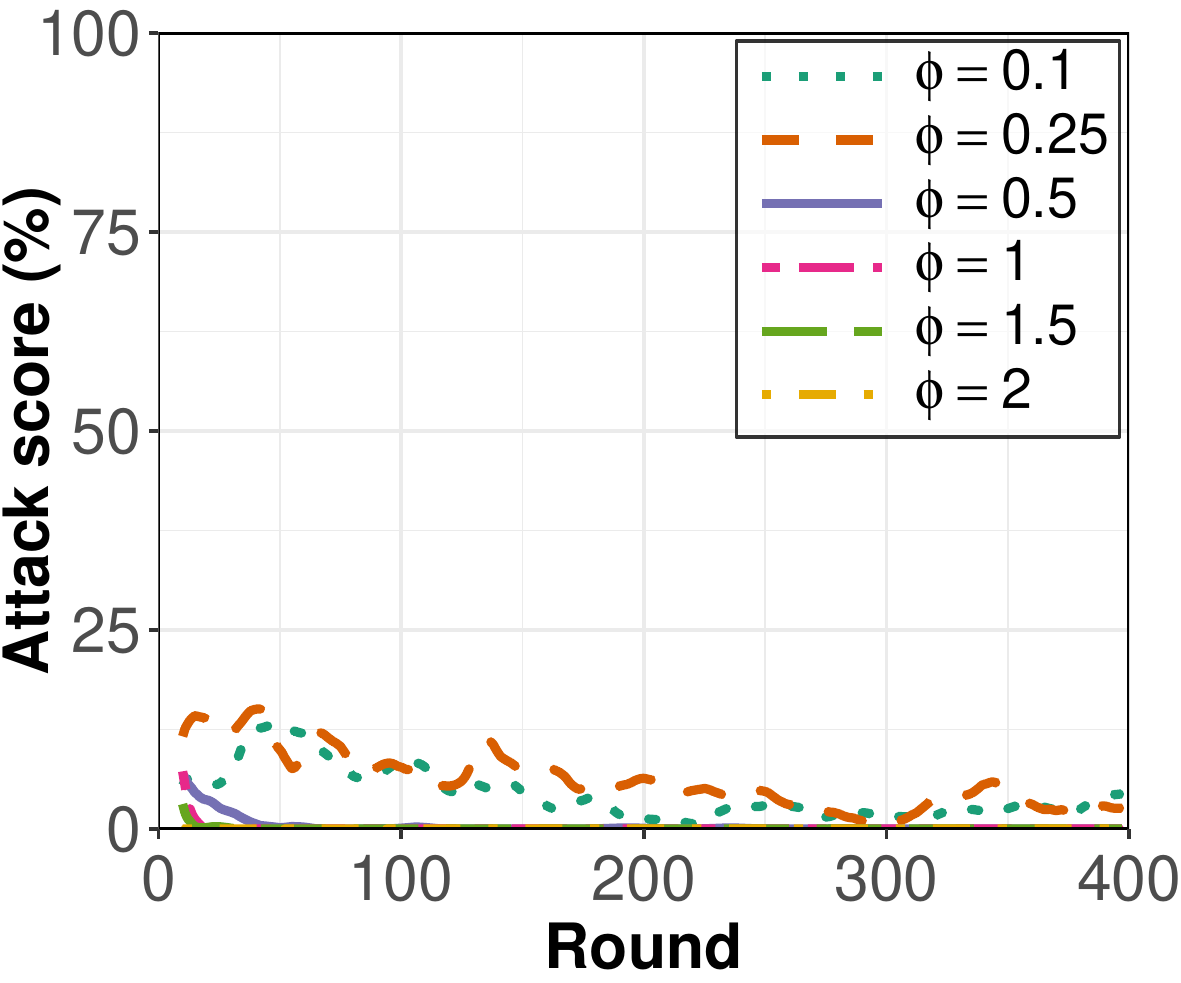}}
    \hfill
  \subfloat[\label{fig:attack-densities-backdoor-accuracy}Accuracy \backdoor{}]{%
        \includegraphics[width=0.5\linewidth]{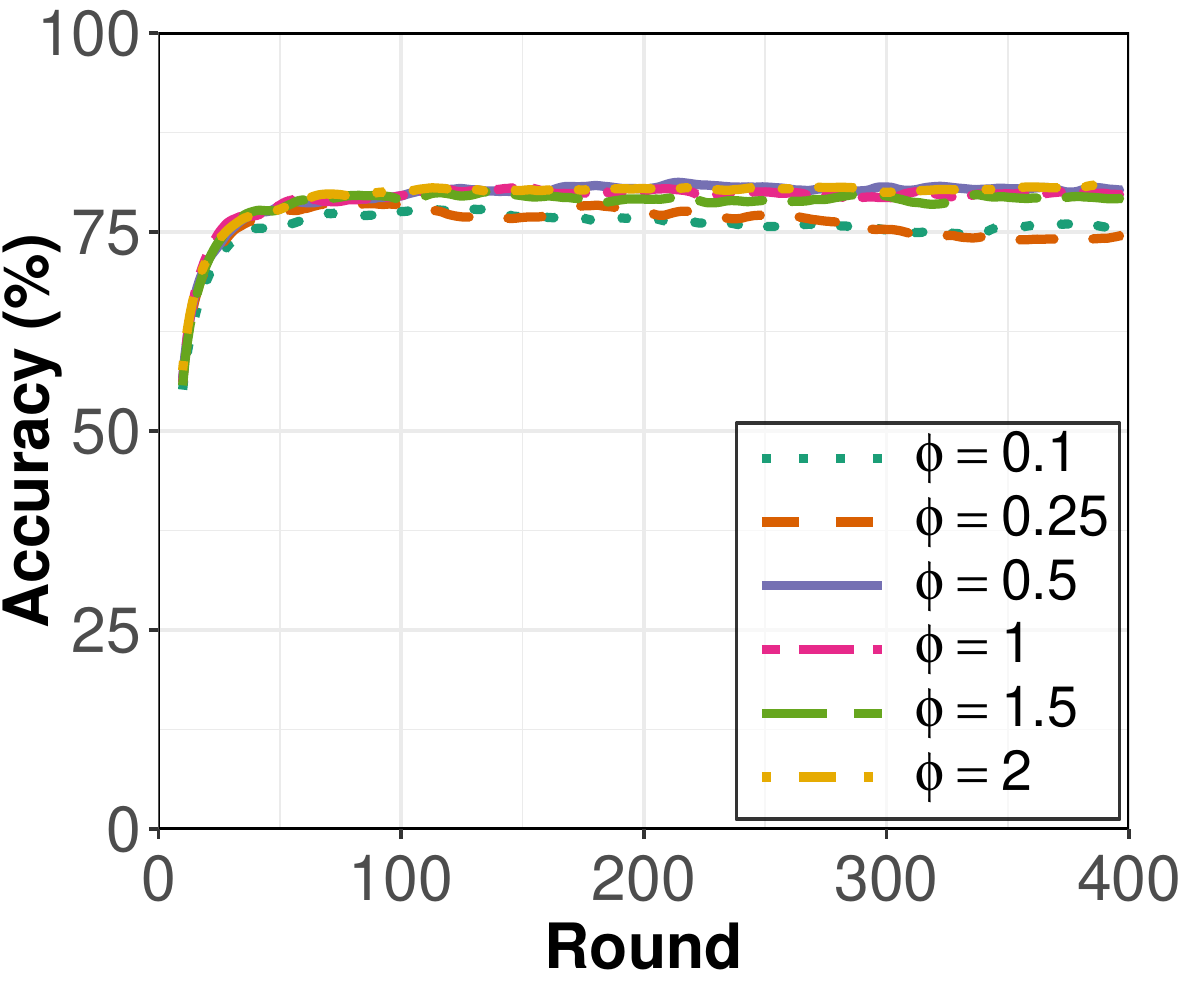}}
    \hfill
  \subfloat[\label{fig:attack-densities-backdoor-attack}Attack score \backdoor{}]{%
        \includegraphics[width=0.5\linewidth]{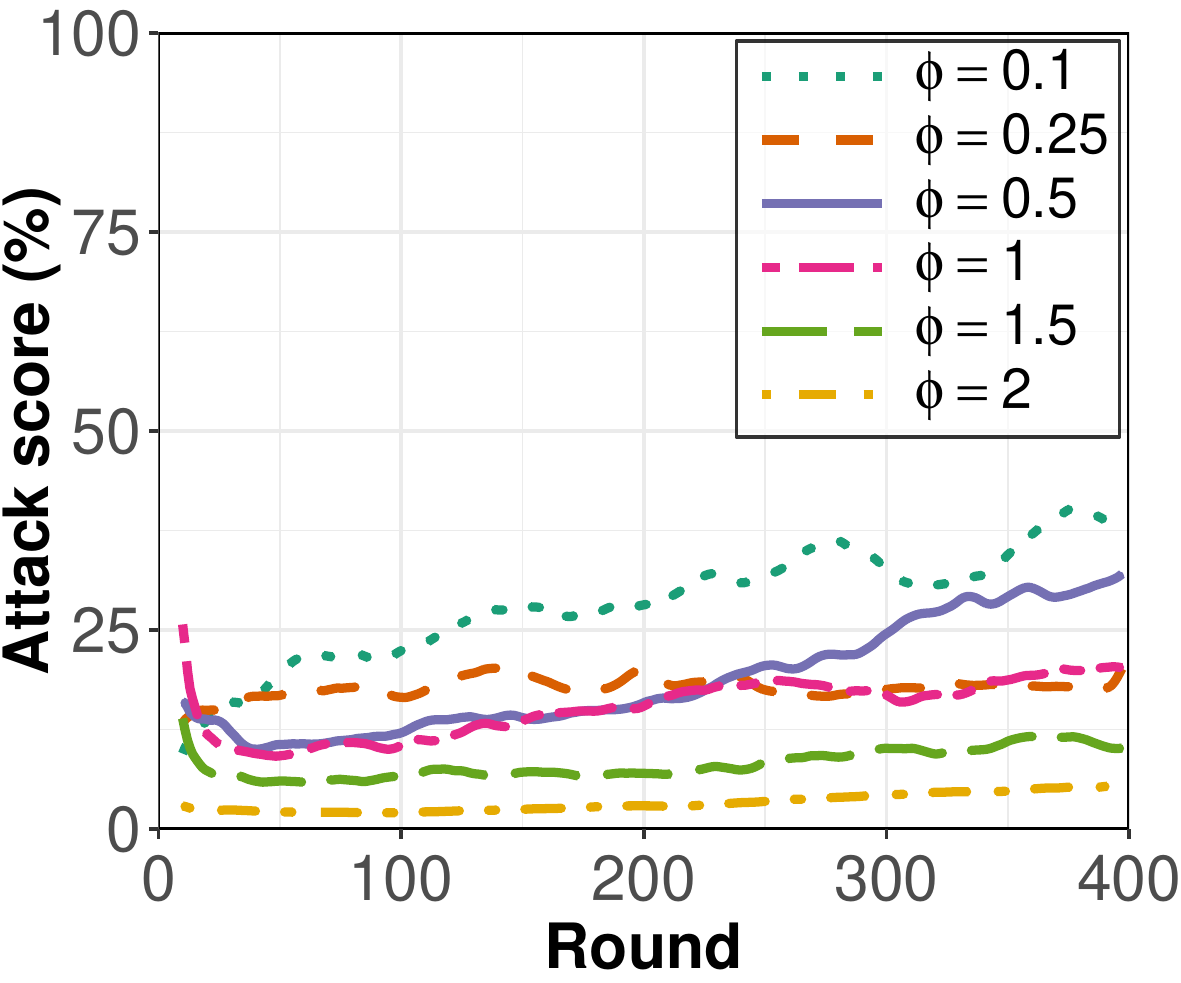}}
    \caption{\label{fig:attack-densities}Accuracy and attack score for the \labelflip{} and \backdoor{} on different attack edge densities. Results generated using the MNIST \cite{mnist} dataset.}
\end{figure}

Figure \ref{fig:attack-densities} illustrates the effect of various attack edge density values on the \labelflip{} and \backdoor{}. It is apparent that the attack edge density has little effect on the converged accuracy (Figures \ref{fig:attack-densities-labelflip-accuracy} and \ref{fig:attack-densities-backdoor-accuracy}).
However, Figure \ref{fig:attack-densities-backdoor-attack} shows how the attack score decreases as the attack edge density increases. This suggests that \algoname{} successfully decreases the utility gained from creating additional Sybils, as it will likely decrease the attack score.
A similar effect can be observed with the \labelflip{} (Figure \ref{fig:attack-densities-labelflip-attack}), where a lower attack edge density results in higher attack score. The attack score of the \labelflip{} also demonstrates how \algoname{} becomes increasingly stronger in reducing the attack score over time. This can be explained by the probabilistic gossiping mechanism causing knowledge about distant nodes to accumulate over time, thereby improving the ability to detect Sybils among direct neighbors.

\subsection{Effect of data distribution}
\label{sec:eval_data_distribution}
\subsubsection{Setup} The method in which the data is distributed over the nodes might influence the attack score and the accuracy of the trained models. To explore this effect, we evaluate \algoname{}'s performance under a variety of data distributions. More specifically, we vary the data distribution between i.i.d. and non-i.i.d. (Dirchlet-based). For the non-i.i.d. scenario, we vary the concentration parameter $\alpha$ within the range $\alpha \in [0.05, 1]$. Furthermore, we fixate the dataset on CIFAR-10.


\subsubsection{Results}
\begin{figure*}[t]
    \centering
    
    \subfloat[\label{fig:data-distributions-labelflip-accuracy}Accuracy \labelflip{}]{
        \includegraphics[width=0.245\linewidth]{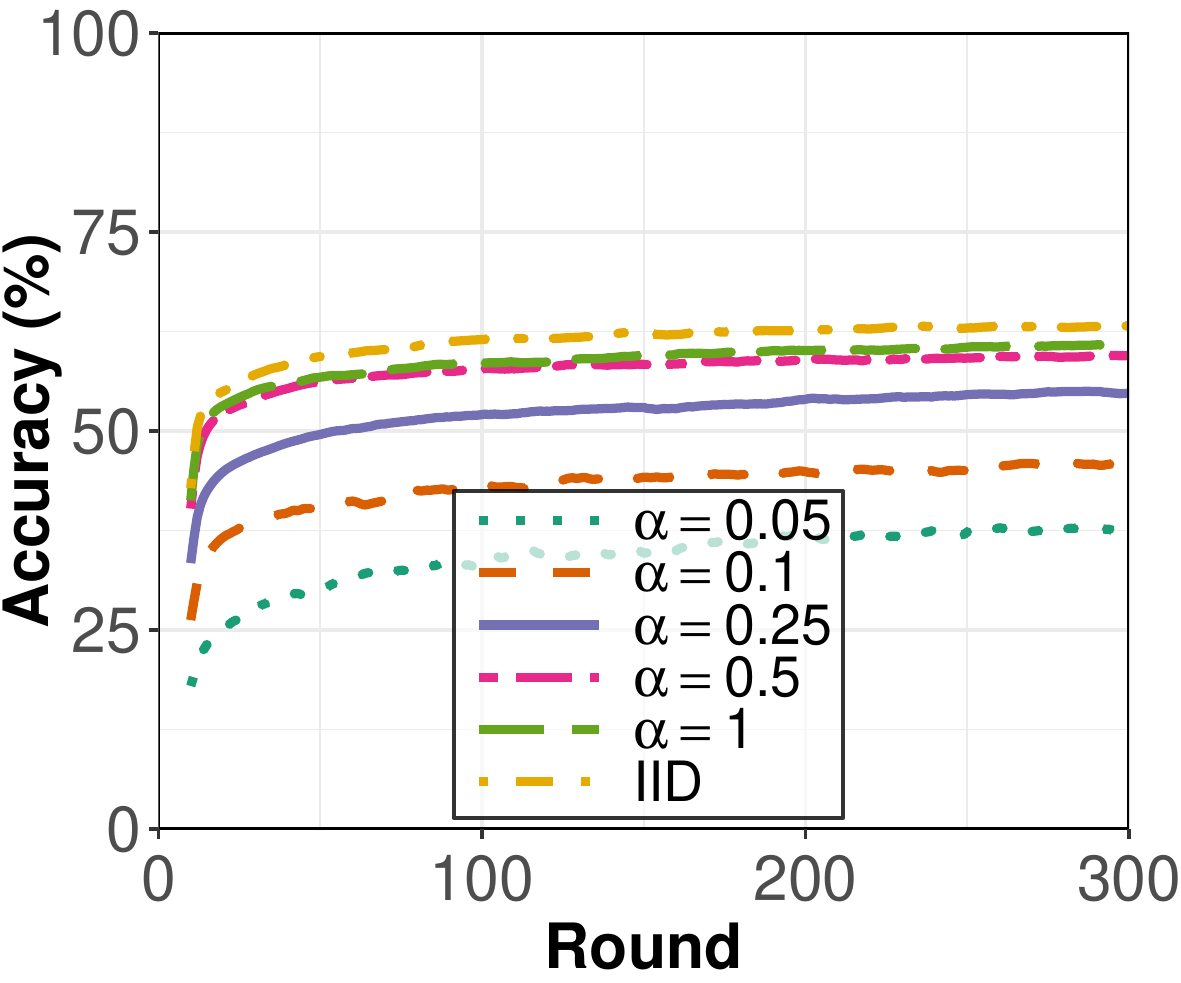}}
  \subfloat[\label{fig:data-distributions-labelflip-attack}Attack score \labelflip{}]{
        \includegraphics[width=0.245\linewidth]{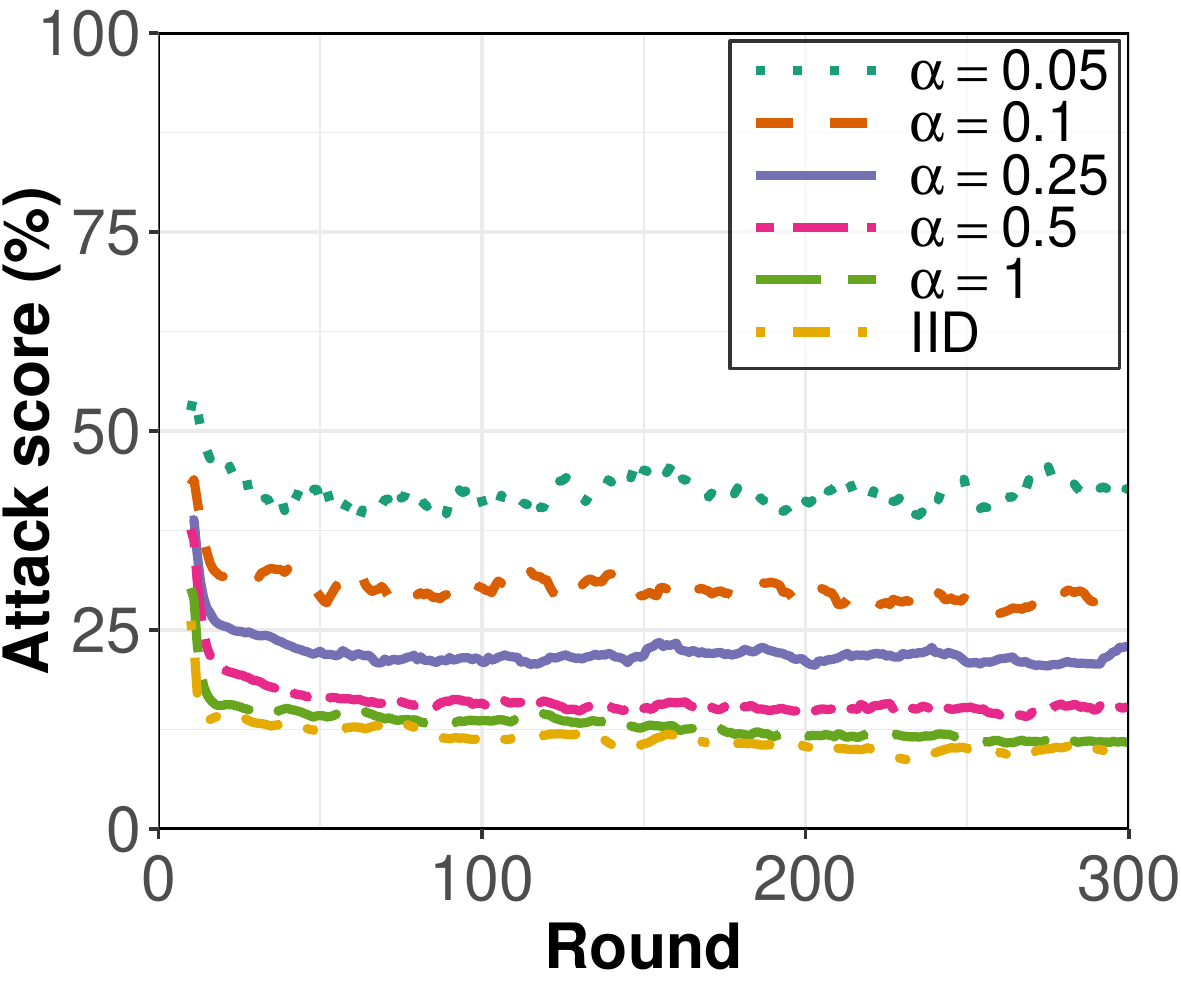}}
  \subfloat[\label{fig:data-distributions-backdoor-accuracy}Accuracy \backdoor{}]{%
        \includegraphics[width=0.245\linewidth]{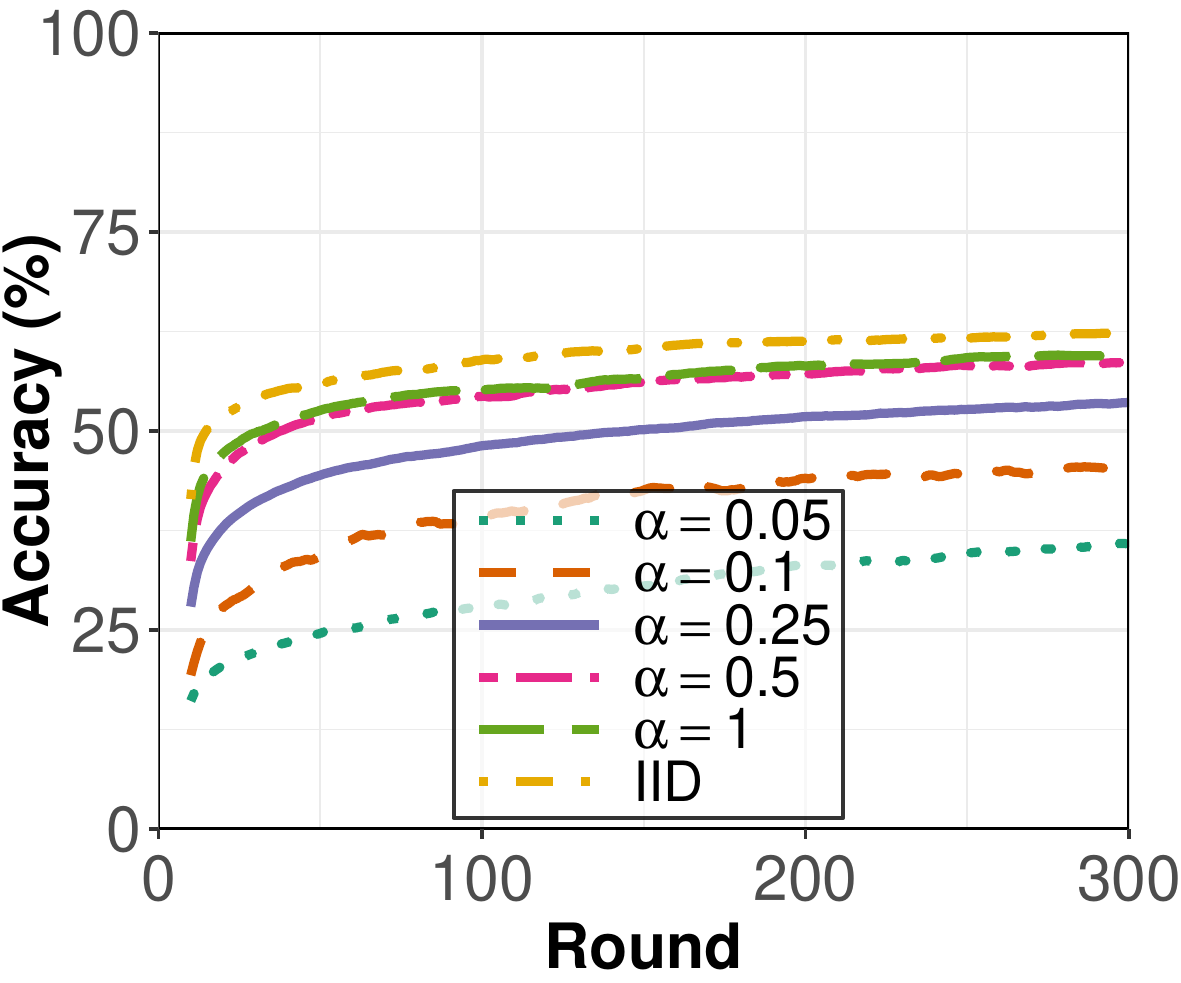}}
  \subfloat[\label{fig:data-distributions-backdoor-attack}Attack score \backdoor{}]{
        \includegraphics[width=0.245\linewidth]{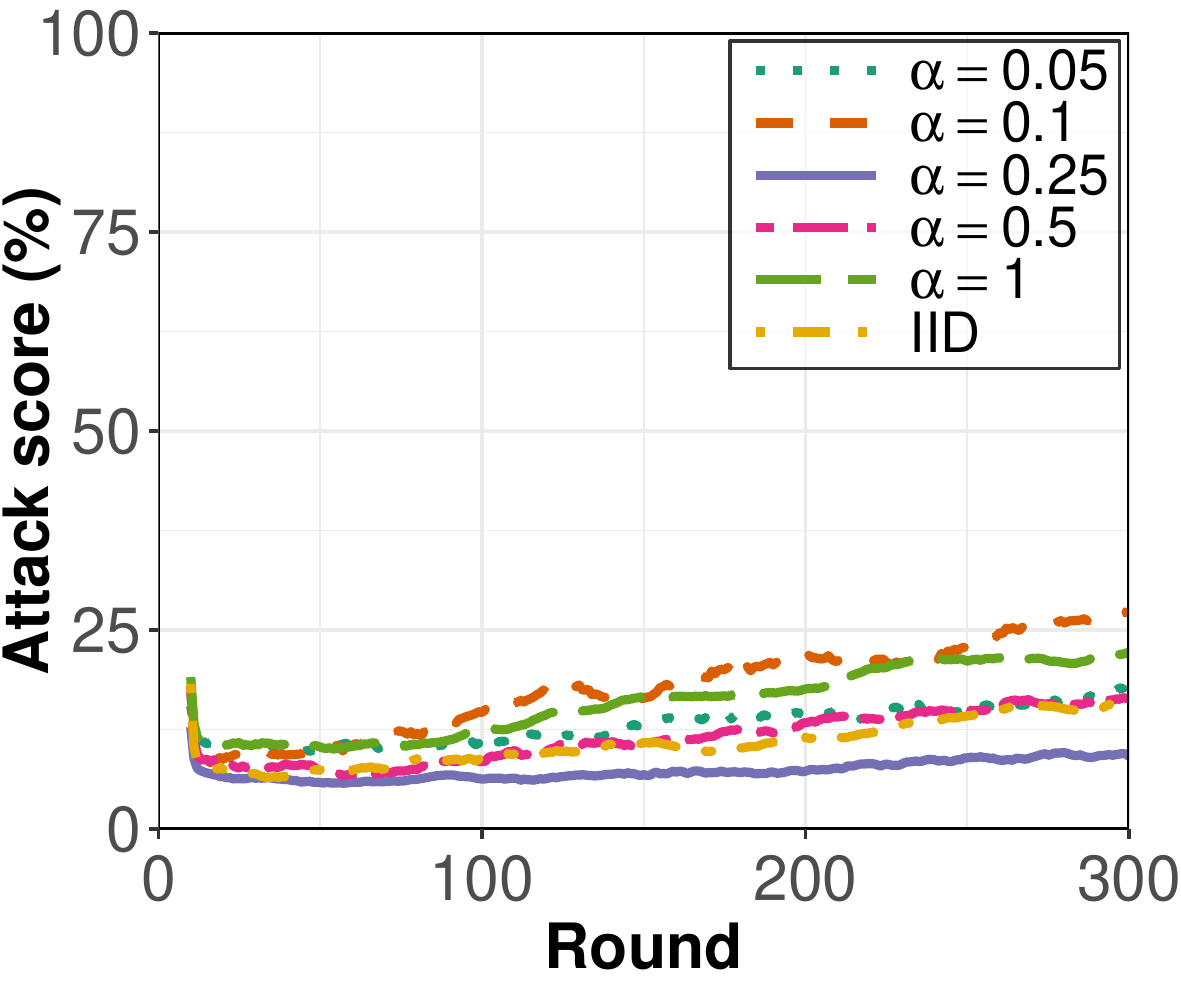}}
    \caption{\label{fig:data-distributions}Accuracy and attack score for the \labelflip{} and \backdoor{} of different data distributions, indicated by the concentration parameter $\alpha$ of the Dirichlet distribution. Results generated using the CIFAR-10 dataset \cite{cifar10}.}
    \subfloat[Accuracy \labelflip{}]{%
       \includegraphics[width=0.245\linewidth]{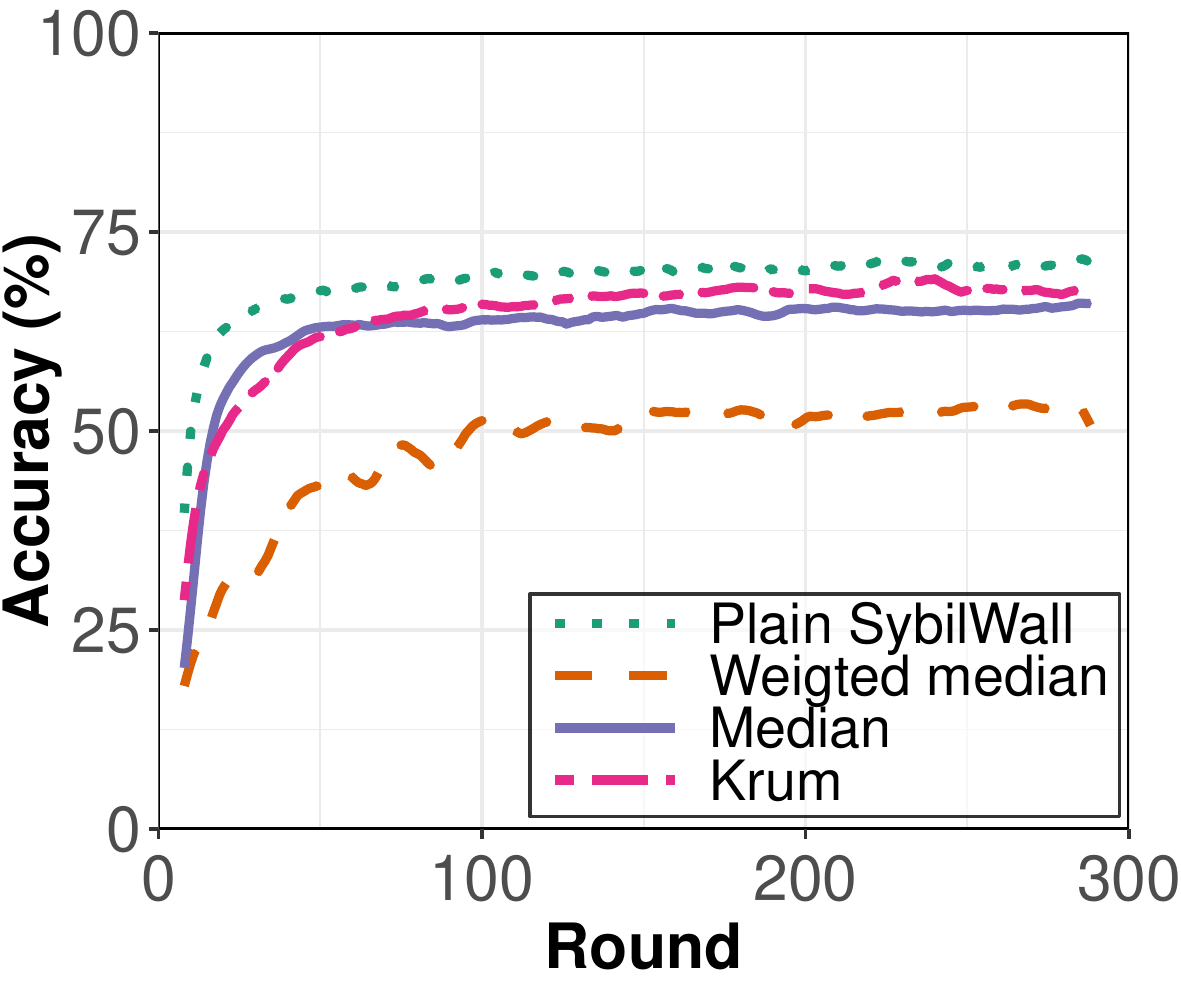}}
  \subfloat[Attack score \labelflip{}]{%
        \includegraphics[width=0.245\linewidth]{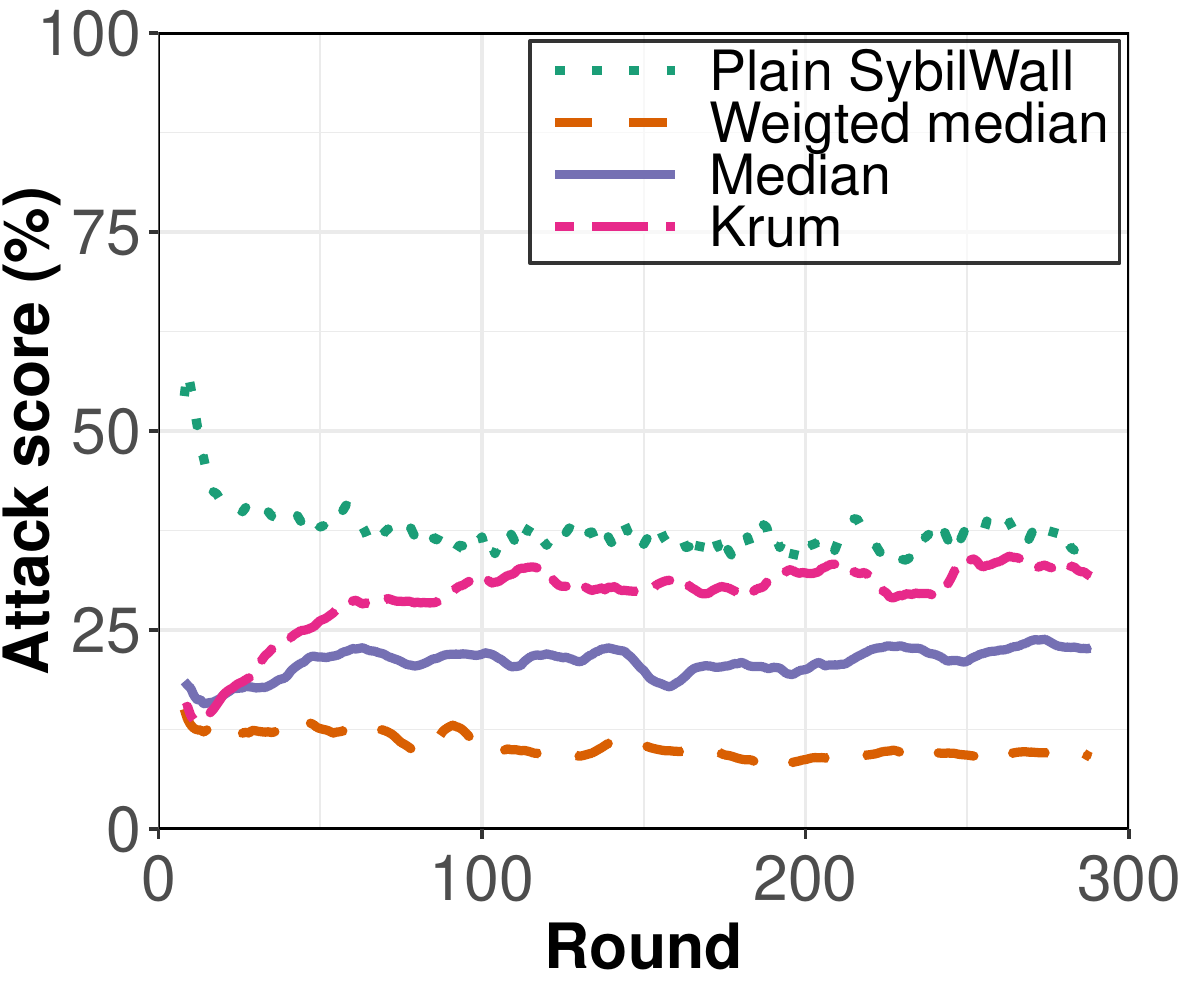}}
        \hfill
  \subfloat[Accuracy \backdoor{}]{%
        \includegraphics[width=0.245\linewidth]{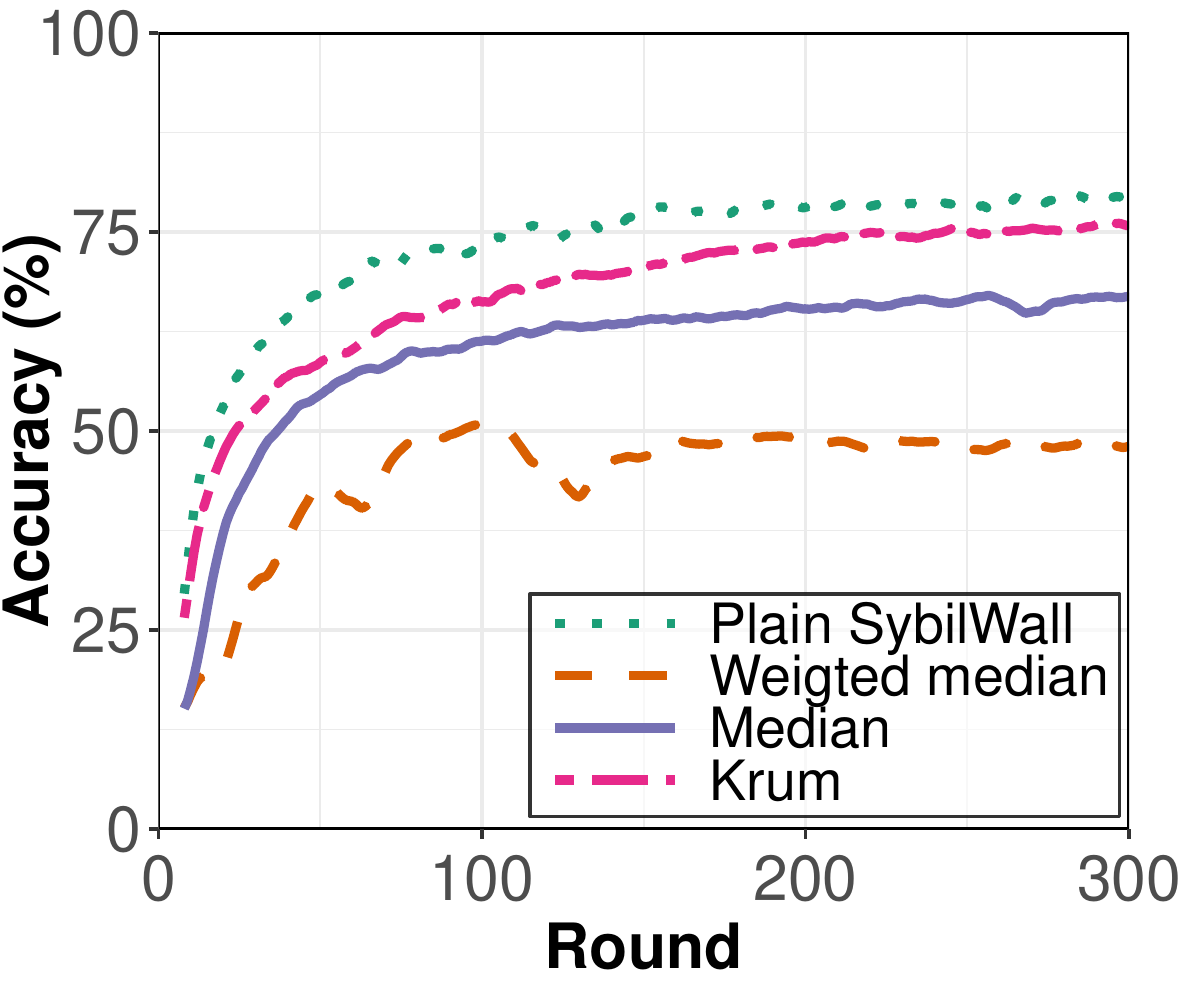}}
  \subfloat[Attack score \backdoor{}]{%
        \includegraphics[width=0.245\linewidth]{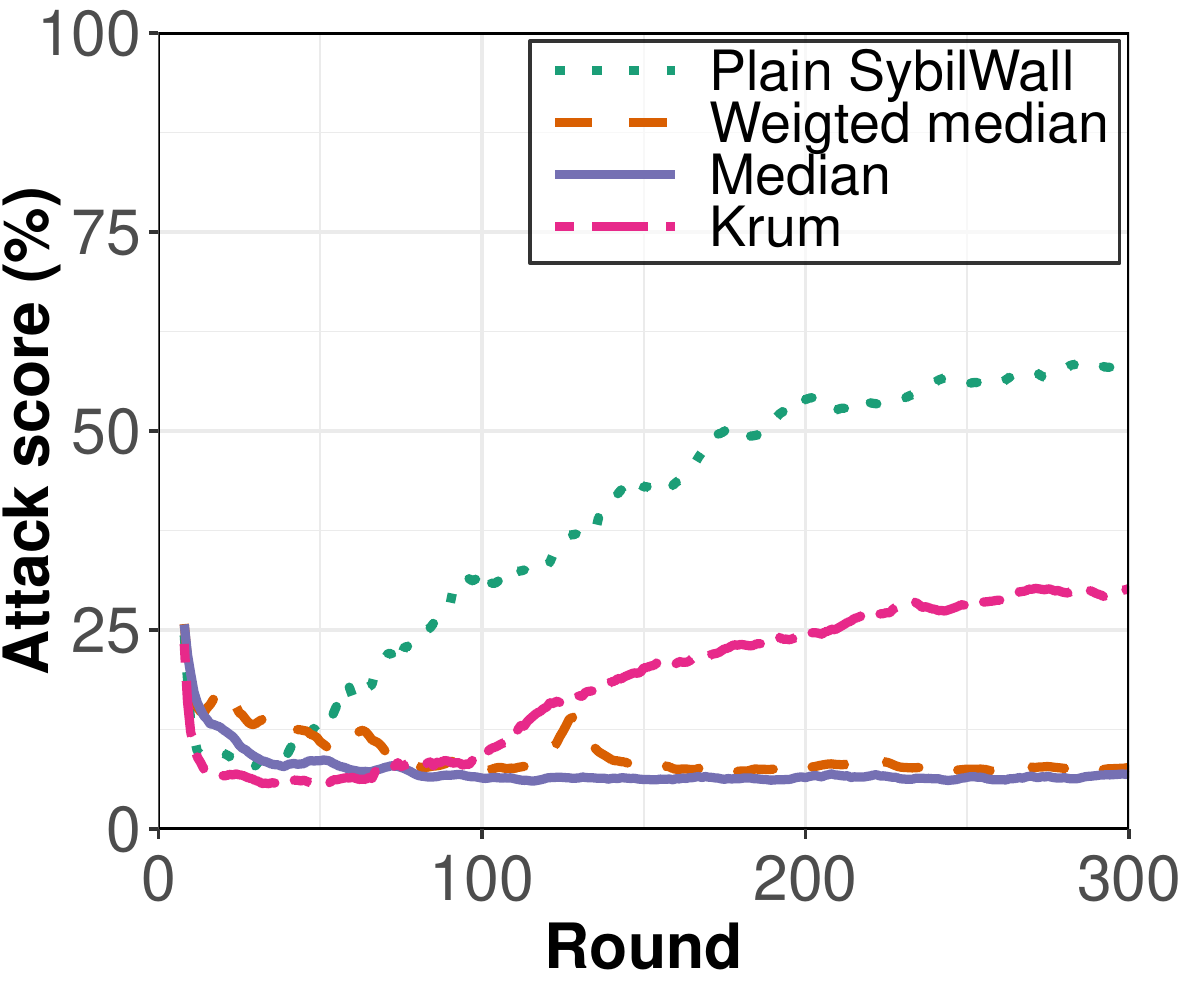}}
  \caption{\label{fig:enhancements}Accuracy and attack score of the \labelflip{} and \backdoor{} for different possible enhancements of \algoname{}. Results generated using the SVHN \cite{svhn} dataset.}
\end{figure*}

Figure \ref{fig:data-distributions} shows the effects of different data distributions on the convergence of the training process. We observe in both the \labelflip{} and \backdoor{} that the accuracy increases as the data is more uniformly distributed (Figures \ref{fig:data-distributions-labelflip-accuracy} and \ref{fig:data-distributions-backdoor-accuracy}). Furthermore, the attack score of the \labelflip{} demonstrates how the attack score decreases as the training data are more uniformly distributed (Figure \ref{fig:data-distributions-labelflip-attack}). This finding suggests that adversaries will be more successful in networks with highly non-i.i.d. data. This is likely due to the decrease in the number of nodes capable of counteracting the \labelflip{}, as they are less likely to possess training samples belonging to the targeted classes.
Lastly, the data distribution does not appear to have a significant effect on the attack score of the \backdoor{}, as no clear trend emerges when varying the data distribution (Figure \ref{fig:data-distributions-backdoor-attack}). This can be explained by the fact that counteracting the backdoor attack does not require the possession of specific training data.

\subsection{Further enhancing \algoname{}}
\label{sec:eval_enhancing}
\subsubsection{Setup}
Given the increasing, although impeded, attack score demonstrated for the \backdoor{} in Section \ref{sec:eval_dataset}, we consider several techniques for additional enhancement of the defensive capabilities of \algoname{}. These augmentations include the following:
\begin{enumerate}[i.]
    \item Median: given the resilience of the Median \cite{median} algorithm in Section \ref{sec:eval_techniques} against attack edge density $\phi = 1$, we implement a combined version of the Median approach and \algoname{}. This is achieved by initially employing \algoname{} to compute a non-normalized aggregation weight in the range $[0, 1]$, followed by the execution of the Median algorithm on the 50\% highest scoring models.
    \item Weighted median: a variant of the Median-based approach, in which scores computed by \algoname{} are adopted as weights for a weighted median aggregation.
    \item Krum-filter: based on the suggestion of prior work \cite{foolsgold}, we combine \algoname{} with Krum, such that the model with the lowest Krum score receives an aggregation weight of 0.
\end{enumerate}
We integrate these augmentations through chaining the aggregation functions, such that the last step of \algoname{}'s aggregation method, a weighted average, is substituted.
We also provide the trends for plain \algoname{} to serve as a baseline. The dataset is fixed to SVHN.

\subsubsection{Results}
Figure \ref{fig:enhancements} illustrates the effect of enhancing \algoname{} with various methodologies. We find that plain \algoname{} achieves the highest accuracy overall, but the worst Sybil resilience.
While each of the evaluated methodologies improves \algoname{}'s defensive capabilities, a trade-off occurs in which accuracy is sacrificed to obtain a decreased attack score. In particular, the low attack score of the weighted median is unmatched, but achieves considerably lower accuracy compared to the alternative methodologies. The Krum-filter-based approach appears to obtain an accuracy comparable to plain \algoname{}, but it obtains the worst Sybil resilience of the evaluated enhancements. However, it exhibits a substantially lower attack score compared to plain \algoname{}. Arguably, the Median-based methodology shows the most promising results, as it achieves to consistently limit the attack score to levels comparable to those of the weighted median methodology, while maintaining a significantly higher accuracy.

\section{Discussion}
During the experimental evaluation of \algoname{}, we performed various experiments to assess the performance and Sybil poisoning resilience of \algoname{}. First, we measured the performance of \algoname{} against 4 widely adopted datasets. We argue that \algoname{} obtained satisfactory accuracy and convergence rate on all of these datasets. Furthermore, the converged accuracy obtained by \algoname{} is similar to that achieved by the FedAvg algorithm in a federated learning setting (Figures \ref{fig:foolsgold-vs-fedavg-cifar-10-fl} and \ref{fig:datasets}). In addition to obtaining satisfactory accuracy on all datasets, we also compared \algoname{} against a number of alternative algorithms and found that \algoname{} was the only evaluated algorithm that consistently scored among the best algorithms in all scenarios. \algoname{} therefore arguably exhibits the overall strongest resilience to Sybil poisoning attacks and possesses the qualities to be considered state-of-the-art. Although the attack score of the \backdoor{} shows a rising trend when employing \algoname{}, we note that the convergence rate is greatly reduced compared to alternative algorithms. This allows honest nodes to stop the training process once the accuracy has converged, thus limiting the success of potential adversaries.

We argue that the aforementioned rising trend demonstrated by the attack score of the \backdoor{} mainly originates from the difficulty of counteracting the effect of the \backdoor{}, as no node will possess training samples directed to mitigate the specific activation pattern. Therefore, \algoname{}'s only method of mitigating the \backdoor{} is by detecting highly similar behavior from Sybils. One might argue that the summation of a node's trained models may not accurately reflect the node's behavioral history, as it is highly affected by the aggregated intermediary model.
Similarly to prior work \cite{foolsgold}, summing a model's trained gradients, rather than the model itself, would arguably improve the representation of a node's behavioral history. Using this approach, a node's history would more closely correspond to a node's training data, thereby decreasing the influence of the aggregated intermediary model. Furthermore, it would also better represent how a node aims to contribute to the aggregated model. 

Assumption \ref{as:no_additional_computation_for_new_sybils} implies that all Sybils will distribute the same model history each training round, as the adversary does not have sufficient computational power to train multiple aggregated intermediary models. This assumption was made to support Assumption \ref{as:sybils_similar_updates}, since Sybils would likely produce highly diverse model histories if each Sybil had a unique aggregated intermediary model. An example of such a situation is click farms \cite{clickfarms}, which provide an adversary with extensive computation capabilities, violating Assumption \ref{as:no_additional_computation_for_new_sybils}. If one were to adopt the use of gradient history rather than model history, the Sybils might exhibit highly similar behavior regardless of the aggregated intermediary model. As Sybils would share the same altered training data, we argue that their summed model gradients will demonstrate more similarity than those of honest nodes. This improvement might eventually lead to the omission of Assumption \ref{as:no_additional_computation_for_new_sybils}.

On the contrary, obtaining a node's model gradients is a non-trivial task in the setting of decentralized learning. It would be highly challenging to verify the validity of the aggregated intermediary model on which the gradients were obtained. This aggregated intermediary model would also need to be communicated to allow neighbors to construct the trained model. One option to verify its validity is by sharing the private training data and violating the user's privacy. However, respecting the user's privacy is one of the fundamental arguments for federated and decentralized learning \cite{federatedlearning}. On the other hand, if one were to allow the usage of unverified aggregated intermediary models, adversaries could trivially launch Sybil poisoning attacks with highly diverse behavior. As an example of such an attack, a Sybil could first train the aggregated intermediary model on the altered training dataset to obtain a malicious trained model $m_s$. Secondly, the Sybil can generate an arbitrary aggregated intermediary model $m_r$ such that the gradients $g$ are defined as $g = m_s - m_r$. Since the validity of the arbitrary aggregated intermediary model $m_r$ cannot be validated, each Sybil could create highly diverse model gradients. This attack is not possible in federated learning, as the aggregated intermediary model is equal for all nodes every round and was created by a central authority.

As previously mentioned, adopting the usage of gradient histories requires the ability to verify the validity of the aggregated intermediary model. Mao et al. \cite{poisonedattackmitigationinDL} suggests a method which achieves this by reaching a global consensus on the aggregated intermediary model. By repeatedly averaging the model with that of neighbors, nodes converge to a globally coherent model under the assumption that every node will participate honestly while attempting to reach consensus.
However, we argue that this assumption does not realistically reflect a deployed decentralized setting and is therefore not applicable to this study. We leave the required analysis for a robust method of verifying the validity of the aggregated intermediary model to future work.

During the evaluation of the effect of varying the attack edge density on the attack score in Section \ref{sec:eval_attack_edge_density}, we found that decreasing the number of Sybils increases the attack score. Therefore, we hypothesize that \algoname{} eliminates the need to amplify a poisoning attack with the Sybil attack, as employing Sybils would result in a lower attack score. However, \algoname{} does not have the ability to successfully mitigate a single-attacker poisoning attack. Mitigating such an attack would require the incorporation of an alternative poisoning attack mitigation algorithm. Section \ref{sec:eval_enhancing} explores further enhancing \algoname{} with a number of such alternative algorithms. Although all enhancements demonstrated increased resilience to Sybil poisoning, they all sacrifice in terms of accuracy. Considering that accuracy is often the primary goal in machine learning \cite{kaur2016survey}, the justification of such enhancements is highly dependent on the application and its users. We leave further enhancing \algoname{} for increased single-attacker poisoning mitigation as a possible research direction for future work. Ideally, this algorithm would increase the resilience against individual attack edges without compromising accuracy.

Furthermore, adversaries may employ a strategy to generate more diverse Sybil model histories. By introducing random noise to the irrelevant weights of the model \cite{foolsgold}, adversaries may be able to significantly increase the diversity among Sybil model histories, resulting in a violation of Assumption \ref{as:sybils_similar_updates}. 
More research is required to accurately filter for relevant weights, which could be achieved through a number of approaches, such as layer-wise relevance propagation \cite{lwrp}, weight magnitude filtering \cite{magnitude}, or empirical weight importance \cite{Luo_2017_ICCV}.

\section{Conclusion}
We have presented \algoname{}, a pioneering algorithm in the mitigation of Sybil poisoning attacks in decentralized learning. Building on the Sybil poisoning mitigation algorithm, FoolsGold \cite{foolsgold} (federated learning), we exploit the increased similarity between the models produced by Sybils over that of honest nodes. We proposed a probabilistic gossiping mechanism to facilitate data dissemination. The disseminated data aids in the mitigation of a poisoning attack amplified by distributing Sybils over the decentralized network. We argue that \algoname{} achieves satisfactory performance on four widely adopted datasets and obtains similar accuracy to federated learning. Furthermore, we empirically evaluated \algoname{} against several alternative algorithms. Our findings indicate \algoname{} to be the only algorithm that consistently scored among the best in all evaluated scenarios, thus arguably outperforming all alternative evaluated algorithms. Although \algoname{} does not fully mitigate targeted poisoning attacks in the form of a \backdoor{}, it manages to greatly decrease the convergence rate of the attacker's success. This enables honest nodes to complete the training process prior to the attack having substantial impact.

We proposed a number of promising future research directions, such as further improving \algoname{} to deflect single attackers, or exploring potential improvements to better mitigate the \backdoor{} by adopting the usage of summed model gradients in the similarity metric.

\bibliographystyle{IEEEtran}
\bibliography{references}

\end{document}